\documentclass[]{aa} 
\usepackage{graphicx}
\usepackage{txfonts}
\usepackage{xcolor}
\usepackage{comment}
\usepackage{hyperref} 
\newcommand{\bl}[1]{\mbox{\boldmath$ #1 $}}
\newcommand{\simname}[1]{\texttt{#1}}
\begin{document} 


   \title{Primordial dust rings, hidden dust mass, and the first generation of planetesimals in gravitationally unstable protoplanetary disks}




   \author{ Eduard I. Vorobyov\inst{1,2},
    Aleksandr M. Skliarevskii\inst{2}, Manuel Guedel\inst{1}, and Tamara Molyarova\inst{2}
          }
   \institute{ University of Vienna, Department of Astrophysics, T\"urkenschanzstrasse 17, 1180, Vienna, Austria; 
   \email{eduard.vorobiev@univie.ac.at} 
   \and
     Research Institute of Physics, Southern Federal University, Rostov-on-Don 344090, Russia   
}

\date{}

   
   \titlerunning{First generation of planetesimals}
   \authorrunning{Vorobyov et al.}

  \abstract
  {}
   {A new mechanism of dust accumulation and planetesimal formation in a gravitationally unstable disk with suppressed magnetorotational instability is studied and compared with the classical dead zone in a layered disk model.}
   {We use numerical hydrodynamics simulations in the thin-disk limit (\simname{FEOSAD} code) to model the formation and long-term evolution of gravitationally unstable disks, including dust dynamics and growth.  }
   {We found that in gravitationally unstable disks with a radially varying strength of gravitational instability a region of low mass and angular momentum transport forms in the inner several astronomical units. This region is characterized by low effective $\alpha_{\rm GI}$ and is similar in characteristics to the dead zone in the layered disk model. As the disk forms and evolves, the GI-induced dead zone accumulates a massive dust ring, which is susceptible to the development of the streaming instability. The model and observationally inferred dust masses and radii may differ significantly in gravitationally unstable disks with massive inner dust rings.}
   {The early occurrence of the GI-induced dust ring followed by the presumed development of the streaming instability suggest that this mechanism may form the first generation of planetesimals in the inner terrestrial zone of the disk. The proposed mechanism, however, crucially depends on the susceptibility of the disk to gravitational instability and requires that the magnetorotational instability be suppressed.}


   \maketitle

\section{Introduction}

Protoplanetary disks form during the gravitational collapse of rotating cloud cores. Both observations and numerical modeling demonstrate that the resulting gas-dust disks can be characterized by a variety of substructures, such as spiral arms, vortices, and clumps \citep{Tobin2016,Huang2018,Varga2021}.  Perhaps the most intriguing among these substructures are rings and gaps, which have been detected via spatially resolved sub-millimeter observations of thermal dust emission and in scattered light in optical/near-infrared wavelengths \citep[e.g.][]{Andrews2018, Long2018, Avenhaus2018,VanDerMarel2019,Zhang2021,Parker2022}

The nature of ring-like structures is not well understood and many theoretical mechanisms have been proposed to explain their origin. Among them are planet-induced rings \citep[e.g.][]{Rice2006,Picogna2015, Dong2015}, which can be interpreted as signposts of planet formation that took place in the disk. Snow lines and dust sintering can also assist in forming dust rings by altering the dust size distribution and the corresponding dust drift velocities \citep{Zhang2015,Okuzumi2016,Pinilla2017}.  Magnetocentrifugal winds can lead to dust accumulation in rings \citep{Riols2020b}. Differential dust drift and/or the back reaction of dust on gas combined with dust growth were also reported to induce pile-up of dust grains in the disk \citep{Drazkowska2016,2017Drazkowska, Gonzalez2017}.
The baroclinic instability induced by dust settling can also act to concentrate dust into rings \citep{Bate2015}.
Globally gravitationally stable disks with enhanced dust-to-gas ratios and low turbulent viscosity can develop dust rings due to the effect known as secular gravitational instability \citep{Takahashi2014}.
Transient dust rings can also form after FU-Orionis-type luminosity bursts and episodes of disk gravitational fragmentation \citep{Vorobyov2020}. We also note that ring structures observed in the dust continuum emission may not be directly related to dust concentration but rather to a peculiar radial dust size distribution in the disk \citep{Akimkin2019}.

Another feasible phenomenon that can assist in dust accumulation and ring formation are dead zones, which are disk regions that are characterized by a reduced rate of mass transport. These features can develop in the regions of magnetized disks where the magnetorotational instability (hereafter, MRI) is suppressed \citep[e.g.,][]{Dzyurkevich2010,Flock2015}. The MRI can provide viscosity via turbulence and the resulting gas surface density profiles of a viscously evolving MRI-active disk is a monotonically declining function of distance from the star \citep{Armitage2022}. However, if a dead zone is present, the gas accumulates in its vicinity due to a reduced rate of mass transport, which in turn leads to the formation of a dust ring in a local pressure maximum \citep[e.g.,][]{Wuensch2005,Pinilla2012,Dullemond2018,Kadam2022}.

Dead zones naturally occur in numerical simulations that consider the ''layered-disk'' model originally proposed in \citet{Gammie1996} and further elaborated in \citet{Armitage2001}. The model suggests that an outer part of the disk is fully MRI-active due to sufficient ionization via cosmic rays penetrating through the entire vertical disk column. These disk regions are MRI-turbulent and the corresponding kinematic viscosity can be characterized by $\alpha_{\rm visc} \approx 10^{-2}$ \citep{Bae2014}, following the turbulent viscosity parametrization of \citet{1973ShakuraSunyaev}. In the inner part of the disk, a~few~$\times$~(0.1--1.0)~au, where the gas density is higher, only the upper disk layers with a column density $\lessapprox 100$~g~cm$^{-2}$ are sufficiently ionized by cosmic rays and hence MRI-active. The rest of the disk vertical column is MRI-dead.  As a result, the effective $\alpha_{\rm visc}$-parameter weighted over the column density of the active and dead layers drops to $\alpha_{\rm visc}< 10^{-3}$, and the mass and angular momentum transport in the inner disk regions is reduced accordingly.  In the innermost disk regions ($< \mathrm{a}~\mathrm{few} \times 10^{-1}$~au), the rising disk temperature and associated thermal ionization of alkaline metals makes the entire vertical column of the disk MRI-active again.
The radial variations in the mass transport efficiency through the inner disk regions lead to a 'traffic jam' situation when gas accumulates near a sharp transition in the $\alpha_{\rm visc}$ value.

Interestingly, dead zones may be a transient phenomenon. Heating of the dead zone owing to residual turbulence and PdV work can raise the gas temperature above 1000~K.  
Thermal ionization of alkaline metals allows fast MRI growth  across most of the dead zone, followed by rapid transport of the inner disk material on to the star, a phenomenon known as an MRI-triggered burst \citep{Armitage2001,2009ZhuHartmannGammie,VorobyovKhaibrakhmanov2020,Kadam2020}. This process can lead to the complete destruction of the dust ring-like structures that have earlier formed in the dead zone. Although the dead zone regenerates soon after the burst, the accumulated dust reservoir is irreversibly lost to the star. This may impede planetesimal formation if the time between outbursts is shorter than the characteristic time of planetesimal formation via the streaming instability \citep{Kadam2022}.

In the recent years, both theoretical models and observational data emerge suggesting that the MRI may be suppressed throughout most of the disk extent and not only in the inner disk regions \citep{Lodato2017, Dullemond2018, Zhang2018, Rosotti2020, DoiKataoka2021, Villenave2022}. In particular, nonideal magnetohydrodynamics effects can suppress the MRI and instead launch magnetocentrifugal winds \citep{BaiStone2013,Gressel2015}. 
The MRI can also be suppressed in the limit of enhanced gravitational instability in the disk \citep{Riols2018}.
Furthermore, observations of  evolved disks in the T~Tauri stage revealed efficient dust settling towards the disk midplane, which would be difficult in the presence of strong MRI-induced turbulence \citep{Rosotti2023},  but see also Sect.~\ref{Sect:caveats} regarding dust settling in gravitationally unstable disks.  In this situation, 
disk magnetocentrifugal winds may act as an alternative mechanism of inward mass transport in the disk,  but their efficiency depends on the poorly constrained disk characteristics, such as magnetic field geometry and the ionization rate \citep{Spruit1996}.  

On the other hand, it is known that young and massive protoplanetary disks can be prone to gravitational instability (hereafter, GI), particularly in the early embedded stage of disk evolution \citep{Kratter2016}. Continual mass loading from the infalling envelope acts to replenish the disk mass loss due to accretion on the star and helps to sustain the disk gravitational instability \citep{2005VorobyovBasu}. As was recently demonstrated by \citet{Vorobyov2023a}, taking disk GI into account has an effect on disk evolution that is similar to the MRI in the layered-disk model. GI has a spatially varying efficiency of mass transport through the disk, being strongest at large radial distances and diminishing in the innermost disk where temperature and sheer are too high for GI to be sustained. The effective $\alpha_{\rm GI}$ parameter, which can be used to describe the efficiency of mass transport if the disk mass is a small fraction of the stellar mass \citep{Vorobyov2010}, has a deep minimum in the innermost disk and is growing further out in the disk.  This may lead to the formation of a dead zone around 1~au, which now has a purely GI origin and is not related to the layered-disk model. Dust that drifts through the disk is efficiently trapped in a local pressure maximum forming at the position of the GI-induced dead zone, provided that the MRI is suppressed and $\alpha_{\rm visc}$ remains low.

In this work, we consider in detail this scenario of the dead zone formation for different model disk realizations. We investigate if the GI-induced dust rings can be favorable sites for planetesimal formation via the streaming instability \citep{Youdin2005,Johansen2011,Yang2017,Umurhan2020}. We also calculate the synthetic observables, such as the intensity of dust radiation at mm-wavebands, and investigate if they can help us to observationally infer the presence of such rings.

The paper is organized as follows. In Sect.~\ref{sec:diskmodel} a description of the numerical model is provided. In Sect.~\ref{Sect:3} the properties of dust rings formed in the layered and GI-controlled disks are analysed. In Sect.~\ref{paramspace} a parameter-space study is conducted. Sect~\ref{Sect:stream-inst} considers the prospects of the streaming instability in our models.
Sect.~\ref{Sect:implications} presents implications for the masses and sizes of dust disks, while in Sect.~\ref{Sect:caveats} we describe the model caveats. Our main conclusions are summarized in Sect.~\ref{Sect:conclude}.

\section{Protostellar disk model}
\label{sec:diskmodel} 
The current work is based on the numerical hydrodynamics simulations that were carried out using the \simname{FEOSAD} code. The numerical model is presented in detail in \cite{2018VorobyovAkimkin}, followed by modifications to account for the adaptive $\alpha$-parameter  \citep{Kadam2019}, updated dust growth scheme \citep{Molyarova2021}, and consideration of the back-reaction of dust onto gas in different drag regimes \citep{Stoyanovskaya2020, Vorobyov2023a}.
Here we only describe the main constituent parts of the numerical model, highlight the details that are relevant for our study, and present the updates applied to the model in addition to those described in the aforementioned papers.

The numerical simulations start from the gravitational collapse of a flattened pre-stellar molecular cloud, followed by the formation of a central protostar and circumstellar disk. The evolution of the disk was computed for about 0.5~Myr after the formation of the protostar.  The equations of hydrodynamics were solved in the thin-disk limit for the gas and dust components of the disk. We used the two-dimensional ($r, \phi$) polar grid extending from 0.2~au to 3500~au. The integration of hydrodynamics equations is carried out using a combination of finite-differences and finite-volume methods with a time-explicit
solution procedure similar in methodology to the ZEUS code \citep{SN1992}. The advection of gas and dust is treated using the third-order-accurate piecewise-parabolic interpolation scheme of
\citet{Colella1984}. The grid contains $400 \times 256$ cells, which are logarithmically spaced in the radial direction and linearly in the azimuthal one. This allows us to treat accurately the processes in the inner disk region, where the numerical resolution reaches $5\times 10^{-3}$~au near the inner computational boundary.  We note that the numerical resolution on the log-spaced grid deteriorates at larger distances but still remains reasonable within 100-200~au, which is the typical extent of the disk in our simulations. In particular, the resolution is $\approx 0.25$~{\color{blue} au} at 10~au and $\approx 2.5$~au at 100~au.  

We note that the adopted thin-disk limit is different from the razor-thin approximation because the vertical scale height of the gas disk is calculated using the assumption of local hydrostatic equilibrium in the gravitational field of both star and disk \citep{VorobyovBasu2009}. This quantity is further used in the calculation of the disk thermal balance by computing the fraction of stellar irradiation absorbed by the disk surface. The stellar mass grows according to the mass accretion rate through the inner computational boundary and the properties of the protostar are calculated using the stellar evolution tracks obtained with the STELLAR code \citep{2008YorkeBodenheimer,2013HosokawaYorke}.

\subsection{\simname{FEOSAD} code: the gaseous component}
\label{sec:gaseous}
The system of equations for the gaseous component consists of the continuity equation, equations describing the gas dynamics, and the energy balance equation. The dynamics of gas is determined by gravity (both central source and disk self-gravity), viscosity, and friction between gas and dust. 
The energy balance in the disk depends on viscous heating, radiative heating (including the radiation of a nascent star and background radiation), radiative cooling, and adiabatic work, which can either heat or cool the local medium. The pertinent equations in the thin-disk limit are as follows.

\begin{equation}
\label{eq:cont}
\frac{{\partial \Sigma_{\rm g} }}{{\partial t}}   + \nabla_p  \cdot 
\left( \Sigma_{\rm g} {\bl v}_p \right) = 0,  
\end{equation}
\begin{eqnarray}
\label{eq:mom}
\frac{\partial}{\partial t} \left( \Sigma_{\rm g} {\bl v}_p \right) +  \nabla_p \cdot \left( \Sigma_{\rm
g} {\bl v}_p \otimes {\bl v}_p \right) & =&   - \nabla_p {\cal P}  + \Sigma_{\rm g} \, {\bl g}_p + \nonumber
\\ 
&+& (\nabla \cdot \mathbf{\Pi})_p  - \Sigma_{\rm d,gr} {\bl f}_p,
\end{eqnarray}
\begin{equation}
\frac{\partial e}{\partial t} +\nabla_p \cdot \left( e {\bl v}_p \right) = -{\cal P} 
(\nabla_p \cdot {\bl v}_{p}) -\Lambda +\Gamma + 
\left(\nabla {\bl v}\right)_{pp^\prime}:\Pi_{pp^\prime}, 
\label{eq:energ}
\end{equation}
where the planar components $(r, \phi)$ are denoted by the subscripts $p$ and $p^\prime$, $\Sigma_{g}$ and $e$ are the gas surface density and the internal energy per surface area, respectively, ${\bl v}_{p}=v_r\hat{{\bl r}}+v_\phi \hat{{\bl \phi}}$ is the gas velocity in the disk plane, $\cal{P}$ is the pressure, integrated in the vertical direction using the ideal equation of state ${\cal P}=(\gamma-1) e$ with $\gamma=7/5$,  ${\bl f}_{p}$  is the drag force per unit mass between gas and dust. 

The gravitational acceleration in the disk plane ${\bl g}_p$ takes into account gas and dust self-gravity in the disk and the gravity of the central star when it is formed. The combined gravitational potential of gas and dust is found by solving the integral form for the potential using the convolution method as laid out in \citet{BT1987}
\begin{eqnarray}
\label{potential}
   \Phi(r,\phi) & =&  \\ \nonumber
   &-& G \int_{\rm r_{\rm sc}}^{r_{\rm out}} r^\prime dr^\prime
                   \int_0^{2\pi}
                \frac{ \left(\Sigma_{\rm g} (r^\prime,\phi^\prime) + \Sigma_{\rm d,tot} (r^\prime,\phi^\prime) \right) d\phi^\prime}
                     {\sqrt{{r^\prime}^2 + r^2 - 2 r r^\prime
                        \cos(\phi^\prime - \phi) }}  \, ,
\end{eqnarray} 
where $r_{\rm sc}$ and $r_{\rm out}$ are the inner and outer extents of the computational domain, $\Sigma_{\rm d,tot}$ is the total mass of dust, and $G$ is the gravitational constant. 
We note that the convolution method does not necessarily require introducing a smoothing length to avoid the singularity when $r=r^\prime$ and $\phi=\phi^\prime$. The details of the smoothing-free method, test problems, and comparison with the method that employs an explicit smoothing term are provided in Appendices~\ref{Sect:gravpot} and \ref{App:test-runs}.

To compute the viscous stress tensor ${\bl \Pi}$, we parameterise the kinematic viscosity owing to the MRI turbulence following \citet{1973ShakuraSunyaev} as
\begin{equation}
\label{eq:nuvisc}
    \nu = \alpha_{\rm visc} c_{\mathrm{s}} H_{\mathrm{g}}, 
\end{equation} where $c_{\mathrm{s}}$ is the sound speed and $H_{\mathrm{g}}$ is the gas vertical scale height. Here, $\alpha_{\rm visc}$ can be either constant in time and space or variable as described in more detail in Sect.~\ref{sec:variable_alpha}. Because the MRI turbulence is likely isotropic, $\alpha_{\rm visc}$ represents not only the efficiency of mass and angular transport in the disk plane but also the efficiency of dust settling in the dust growth model described in Sect.~\ref{sect:dustgrowth}.  Radiative cooling  and  heating are denoted by $\Lambda$ and $\Gamma$, respectively. The latter depends on the irradiation temperature at the disk surface $T_{\rm irr}$ accounting for stellar and background blackbody irradiation, for the exact expressions see \citet{2018VorobyovAkimkin}. We set the background temperature $T_{\rm b.g.} = 15$~K.

\subsection{\simname{FEOSAD} code: the dust component}
\label{sect:dustgrowth}
\begin{figure*}
\begin{centering}
\includegraphics[width=\textwidth]{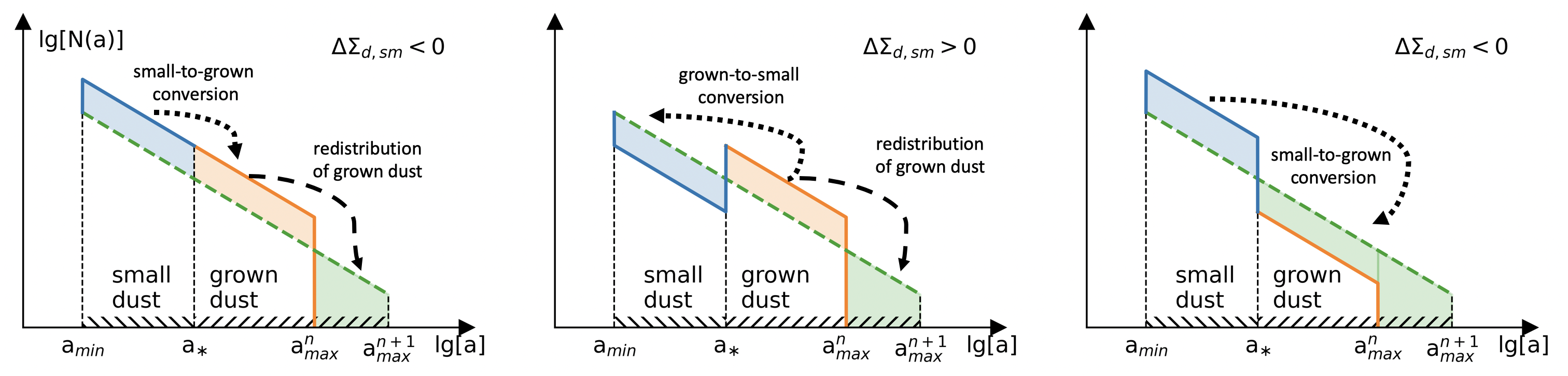} 
\par \end{centering}
\caption{Graphical representation of the conversion between grown and small dust shown for the case of $a_{\rm max}^{\rm n+1} > a_{\rm max} ^{n}$. The solid blue and orange lines indicate the dust distribution at the current time step $n$. The next time step $n+1$ is denoted by the green dashed line. Three cases are presented in the three panels from left to right: (i) the case with a continuous distribution at $a_\ast$; (ii) discontinuous distribution with a dominance of grown dust turning into a continuous distribution, (iii) redistribution of the excess of small dust to provide a continuous distribution.}
\label{fig:scheme}
\end{figure*}
The dust component is divided into two populations: (i) small dust, which are grains with a size\footnote{Here and further in the text by the size of dust grains we mean its radius.} between $a_{\rm min}=5\times 10^{-3} \ \mu \rm m$  and $a_{*} = 1 \ \mu \rm m$ and (ii) grown dust ranging in size from $a_{*}$ to a maximum $a_{\rm max}$, the value of which is variable in space and time. Initially, all dust in the collapsing prestellar cloud is in the small dust population. Small dust  can grow and turn into grown dust as the disk forms and evolves. It is assumed that dust in both populations is distributed over size according to a simple power law: 
\begin{equation}
N(a) = C \cdot a^{ - {\rm p}},
\label{eq:dustdistlaw}
\end{equation} where $N(a)$ is the number of dust particles per unit dust size, $C$ is a normalization constant, and ${\rm p} = 3.5$ (not to be confused with $p$ as a planar component index in Equations~\eqref{eq:cont}--\eqref{eq:momDlarge}). We note that the power index $\rm p$ is kept constant during the considered disk evolution period. A more sophisticated approach requires solving for the Smoluchowski equation for multiple dust bins and is beyond the scope of the present study.

We solve the continuity equations separately for the grown and small dust ensembles. However, the momentum equation is solved only for the grown dust, because small dust is assumed to be dynamically linked to the gas. The system of hydrodynamics equations for dust in the zero-pressure limit is written as:
\begin{equation}
\label{contDsmall}
\frac{{\partial \Sigma_{\rm d,sm} }}{{\partial t}}  + \nabla_p  \cdot 
\left( \Sigma_{\rm d,sm} {\bl v}_p \right) = - S(a_{\rm max}),  
\end{equation}
\begin{equation}
\label{contDlarge}
\frac{{\partial \Sigma_{\rm d,gr} }}
{{\partial t}}  + \nabla_p  \cdot 
\left( \Sigma_{\rm d,gr} {\bl u}_p \right) =  \nabla \cdot \left( D \Sigma_{\rm g} \nabla \left( {\Sigma_{\rm d,gr} \over \Sigma_{\rm g}} \right)  \right) +
S(a_{\rm max}),
\end{equation}
\begin{eqnarray}
\label{eq:momDlarge}
\frac{\partial}{\partial t} \left( \Sigma_{\rm d,gr} {\bl u}_p \right) +  [\nabla \cdot \left( \Sigma_{\rm d,gr} {\bl u}_p \otimes {\bl u}_p \right)]_p  &=&   \Sigma_{\rm d,gr} \, {\bl g}_p + \nonumber \\
 + \Sigma_{\rm d,gr} \, {\bl f}_p + S(a_{\rm max}) {\bl v}_p,
\end{eqnarray}
where $\Sigma_{\rm d,sm}$ and $\Sigma_{\rm d,gr}$ are the surface densities of small and grown dust, respectively, and ${\bl u_{p}}$ are the planar components of the grown dust velocity.  Here, $D$ is the turbulent diffusivity of grown dust, which is related to the kinematic viscosity as $D=\nu / \mathrm{Sc}$ \citep{1988ClarkePringle}. The Schmidt number $\mathrm{Sc}$ is taken to be unity in this study.  We note that in the continuity equation for small dust the velocity of gas ${\bl v_{p}}$ is used because small dust is strictly linked to gas. 
We provide the justification on the applicability of the hydrodynamics equations to describing dust dynamics and on the assumption of coupled dynamics of small dust to gas in \cite{Vorobyov2022}.

The grown dust dynamics is sensitive to the properties of surrounding gas. The drag force (per unit mass) links dust with gas and can be written as \citep{Weidenschilling1977}:
\begin{equation}
    {{\bl f}} = \dfrac{1} {2 m_{\rm d}} C_{\rm D} \, \sigma \rho_{\rm g} ({{\bl v_p}} - {{\bl u_p}}) |{{\bl v_p}} - {{\bl u_p}}|,
\end{equation}
where $\sigma$ is the dust grain cross section, $\rho_{\rm g}$ the volume density of gas, $m_{\rm d}$ the mass of a dust grain, and $C_{\rm D}$ the dimensionless friction parameter. The latter is described in details in \citet{Vorobyov2023a} and is based on the works of \citet{Henderson1976} and \citet{Stoyanovskaya2020}.
The use of the Henderson friction coefficient allows us to treat the drag force in two different regimes, depending on the local conditions and dust properties. More specifically, we consider the Epstein regime, and Stokes linear and non-linear regimes. To account for the back-reaction of grown dust on dust, the term $\Sigma_{\rm d,gr} f$ is symmetrically included in both the gas and dust momentum equations. 

Since grown dust in our model has a spectrum of sizes from $a_\ast$ to $a_{\rm max}$, the values of $\sigma$ and $m_{\rm d}$ have to be weighted over this spectrum. We note that the span between $a_\ast$ to $a_{\rm max}$ may become as large as  several orders of magnitude during the disk evolution. Since we set $p=3.5$, small grains near $a_\ast$ will dominate  the value of $\sigma$, while large grains near $a_{\rm max}$ will mostly determine the value of $m_{\rm d}$.   On the other hand, we are interested in the dynamics of dust grains that are the main mass carriers. Therefore, we use the maximum size of dust grains $a_{\rm max}$ when calculating the values of $\sigma$ and $m_{\rm d}$. The friction force $f$ thus derived would describe the dynamics of the main dust mass carriers. A more consistent approach requires introducing multiple bins for the entire size spectrum of grown dust and is outside the scope of the current work.


The term $S(a_{\rm max})$ that enters the equations for the dust component is the conversion rate between small and grown dust populations.  We assumed that the distribution of dust particles over size follows the form given by Equation~\eqref{eq:dustdistlaw} for both small and grown populations. 
Furthermore, the distribution is assumed to be continuous at $a_{*}$. 
Our scheme is constructed so as to preserve continuity at $a_{*}$ by writing the conversion rate of small to grown dust in the following form:
\begin{equation}
    S(a_{\rm max}) = - \dfrac{\Delta \Sigma_{\rm d,sm}} {\Delta t},
\end{equation}
where
\begin{equation}
\label{final}
    \Delta\Sigma_{\mathrm{d,sm}} = \Sigma_{\mathrm{d,sm}}^{n+1}- \Sigma_{\mathrm{d,sm}}^{n} =
    \frac
    {
    \Sigma_{\rm d,gr}^n \int_{a_{\rm min}}^{a_*} a^{3-\mathrm{p}}da - 
    \Sigma_{\rm d,sm}^n \int_{a_*}^{a_{\mathrm{max}}^{\rm n+1}} a^{3-\mathrm{p}}da
    }
    {
    \int_{a_{\rm min}}^{a_{\mathrm{max}}^{n+1}} a^{3-\mathrm{p}}da
    },
\end{equation}
where indices $n$ and $n+1$ denote the current and next hydrodynamic steps of integration, respectively, and $\Delta t$ is the hydrodynamic time step. The adopted scheme effectively assumes that dust growth smooths out any discontinuity in the
dust size distribution at $a_\ast$ that may appear due to differential drift of small and grown dust populations. The conversion process between small and grown dust populations is schematically illustrated in Figure~\ref{fig:scheme}. A more detailed description of the scheme is presented in~\citet{Molyarova2021} and \citet{Vorobyov2022}.

The value of $S(a_{\rm max})$ depends only on the local maximal size of dust $a_{\rm max}$, since the values of $a_{\rm min}$ and $a_\ast$  are fixed in our model. In particular, $a_{\rm max}$ is not a constant of space and time but is evolving with the disk. At the beginning of the simulations all grains are in the form of small dust, namely, $a_{\rm max} = 1.0$~$\mu$m in the collapsing prestellar core. During the disk formation and evolution process the maximal size of dust particles usually increases. The change in $a_{\rm max}$ within a particular numerical cell can occur due to collisional growth or via advection of dust through the cell. The equation describing the dynamical evolution of $a_{\rm max}$ is as follows:
\begin{equation}
{\partial a_{\rm max} \over \partial t} + ({\bl u}_{p} \cdot \nabla_p ) a_{\rm max} = \cal{D},
\label{eq:dustA}
\end{equation}
where the rate of dust growth due to collisions and coagulation is computed in the monodisperse approximation \citep{Birnstiel2012}
\begin{equation}
\cal{D} = {\rho_{\rm d} \mathit{u}_{\rm rel} \over \rho_{\rm s}}.
\end{equation}
This rate includes the total volume density of dust $\rho_{\rm d}$, the dust material  density $\rho_{\rm s} = 2.24$~g~cm$^{-3}$ \citep{Draine2001}, and the relative velocity of particle-to-particle collisions defined as $\mathit{u}_{\rm rel} = (\mathit{u}_{\rm th}^2 + \mathit{u}_{\rm turb}^2)^{1/2}$, where $\mathit{u}_{\rm th}$ and $\mathit{u}_{\rm turb}$ account for the Brownian and turbulence-induced local motion, respectively. When calculating the volume density of dust, we take into account dust settling by calculating the effective scale height of grown dust  $H_{\rm d}$ via the corresponding gas scale height $H_{\rm g}$, $\alpha_{\rm visc}$ parameter, and the Stokes number as
\begin{equation}
    H_{\rm d} = H_{\rm g} \sqrt{ {\alpha_{\rm visc} \over \alpha_{\rm visc}+\mathrm{St} } }.
    \label{eq:dust-scale-height}
\end{equation}

Dust growth in our model is limited by collisional fragmentation and drift. 
We take into account the fragmentation barrier by calculating the characteristic fragmentation size as \citep{2016Birnstiel}:
\begin{equation}
    a_{\rm frag}=\frac{2\Sigma_{\rm g} \mathit{u}_{\rm frag}^2}{3\pi\rho_{\rm s} \alpha_{\rm visc} c_{\rm s}^2},
\label{eq:afrag}
\end{equation}
where $\mathit{u}_{\rm frag}$ is the fragmentation velocity, namely, a threshold value of the relative velocity of dust particles at which collisions result in fragmentation rather than coagulation. In the current study, we adopt $\mathit{u}_{\rm frag} = 3$~m~s$^{-1}$ \citep{Blum2018}. If $a_{\rm max}$ becomes greater than $a_{\rm frag}$, we stop the growth of dust and set $a_{\rm max} = a_{\rm frag}$.
We note that if the fragmentation barrier is reached and dust growth halts ($a_{\rm max}=a_{\rm frag}$), the local conditions in the disk can change such that the value of fragmentation barrier decreases (for instance, if the gas density decreases or temperature rises). If this occurs, we reduce $a_{\rm max}$ to adjust it to the new value of $a_{\rm frag}$. We note that the so-called drift barrier is accounted for self-consistently via the computation of the grown dust dynamics.


\subsection{Viscosity model}
\label{sec:variable_alpha}
The hydrodynamic model includes the treatment of turbulent viscosity according to the approach of \citet{1973ShakuraSunyaev}. The viscosity is parametrized by the $\alpha_{\rm visc}$-parameter, which can be either constant in space and time or adaptive. The latter case is implemented using the concept of a ``layered'' disk \citep{Gammie1996, Armitage2001}. The details on the implementation are presented in \citet{Kadam2019} based on the work of \citet{Bae2014}.
In particular, the model assumes that a surface layer with column density $\Sigma_{\rm a}$ is sufficiently ionized by cosmic rays to be MRI active. If the local gas surface density of the disk $\Sigma_{\rm g}$ is lower than $2\times \Sigma_{\rm a}$, the entire vertical column of the disk is MRI active. 
In the opposite case, a region below the MRI-active layer exists where the MRI is suppressed. The mathematical expression for $\alpha_{\rm visc}$ in this model is written following \cite{Bae2014} as:
\begin{equation}
    \alpha_{\rm visc} = \dfrac{\Sigma_{\rm a} \alpha_{\rm a} + \Sigma_{\rm d} \alpha_{\rm d}}{0.5 \times \Sigma_{\rm g}},
    \label{alpha:visc}
\end{equation}
where $\Sigma_{\rm d}$ is the thickness of the MRI-dead layers and $\Sigma_{\rm g} = \Sigma_{\rm a} + \Sigma_{\rm d}$ is the total surface density of gas. 
A factor of $0.5$ appears in the denominator due to the fact that $\Sigma_{\rm a}$ is the thickness of the MRI-active layer from the disk surface to the disk midplane and $\Sigma_{\rm g}$ is the total gas surface density from the upper to the lower disk surface.
The quantities $\alpha_{\rm a}$ and $\alpha_{\rm d}$ are the viscosity parameters applied to the MRI-active and MRI-dead layers of the disk, respectively.
In this study, the thickness of the active layer is set equal to $\Sigma_{\rm a} = 100$~g~cm$^{-2}$ and the corresponding $\alpha_{\rm a} = 10^{-2}$. 
In the MRI-dead layer the viscosity parameter is set equal to  $\alpha_{\rm d} = 10^{-5}$, reflecting the fact that the MRI-dead layer is likely to have some nonzero residual transport.

\subsection{Initial and boundary conditions}
Simulations start from the gravitational collapse of a flattened prestellar core, consisting of gas and small dust.  As the core contracts gravitationally, it spins up and a centrifugal disk forms when the in-spiralling gas hits the centrifugal barrier near the stellar surface. In our case, because of the use of the sink cell, this would be the inner computational boundary at $r_{\rm sc}=0.2$~au.  Subsequently, the disk grows in size and mass owing to infall from progressively outer layers of the contracting cloud, while the central star gains mass via accretion through the inner computational boundary. Because of the adopted thin-disk limit, the matter from the contracting core lands on the the disk outer edge but this is a reasonable approximation for a collapsing cloud \citep{Visser2009}.
The initial mass of the core in the fiducial model is $M_{\rm core} = 0.53 M_{\odot}$. The core rotation is determined by setting the ratio of rotational-to-gravitational energy $\beta = 2.3 \times 10^{-3}$. The value is within the limits inferred from prestellar cloud cores \citep{Caselli2002}.

Initially, the gas surface density and angular velocity of the natal prestellar core are distributed as follows \citep{1997Basu}:
\begin{equation}
    \Sigma_{\rm g}(r)=\frac{r_0\Sigma_{\rm 0,g}}{\sqrt{r^2+r_0^2}},
\label{eq:coredens}
\end{equation}
\begin{equation}
    \Omega_{\rm g}(r)=2\Omega_{\rm 0,g}\bigg(\frac{r_0}{r}\bigg)^2\left[\sqrt{1+\left(\frac{r}{r_0}\right)^2}-1\right],
\label{eq:corevel}
\end{equation}
where $\Sigma_{\rm 0,g} = 0.385$~g~cm$^{-2}$ is the surface density  and $\Omega_{\rm 0,g} = 5.1$~km~s$^{-1}$~pc$^{-1}$ is the angular velocity, both defined at the core centre. The radius of the near-uniform region in the centre of the core is $r_{0} = 617.2$~au. The total dust-to-gas mass ratio $\xi_{\rm d2g}=\Sigma_{\rm d,tot}/\Sigma_{\rm g}$ is set equal to the interstellar medium value 0.01. 
The initial values of the small and grown dust surface densities are $\Sigma_{\rm d,sm}(r) = 0.01 \times \Sigma_{\rm g}$ and $\Sigma_{\rm d,gr}(r)=0$, respectively. The core and the subsequently formed disk are heated by the background radiation with a temperature of $T_{\rm bg}= 20$~K, also adopted as the cloud's initial temperature. We emphasize that the disk evolution resulting from the collapse of prestellar cores in our models is weakly sensitive to the particular choice of the initial surface density and angular velocity radial distributions  for as long as $M_{\rm core}$ and $\beta$ of the prestellar cores are similar \citep{2012ARep...56..179V}.

The innermost disk region between the inner disk edge at $r_{\rm sc}=0.2$~au and the star is replaced with a sink cell, which ensures a free mass exchange (inflow and outflow) across the sink-disk interface \citep[see][for details]{2018VorobyovAkimkin}.  We emphasize that the size of the sink cell in our simulations is notably smaller than in many other global disk simulations over timescales of hundreds of thousands of years.
The outer boundary condition allows free mass outflow, but mass inflow from outside the computational domain is prohibited.

\section{Primordial rings of viscous and gravitational origin}
\label{Sect:3}
In this section, we consider the formation of dust rings in the layered-disk model, which is characterized by a radially varying $\alpha_{\rm visc}$-parameter.  We also compare the dust rings in the layered disk model with those formed in a GI-controlled disk, in which $\alpha_{\rm visc}$ is a constant of time and space and is set equal to a small value of $10^{-4}$. In both cases, disk self-gravity is considered and it plays a dominant role in the GI-controlled model.

\begin{figure}
\begin{centering}
\includegraphics[width=\linewidth]{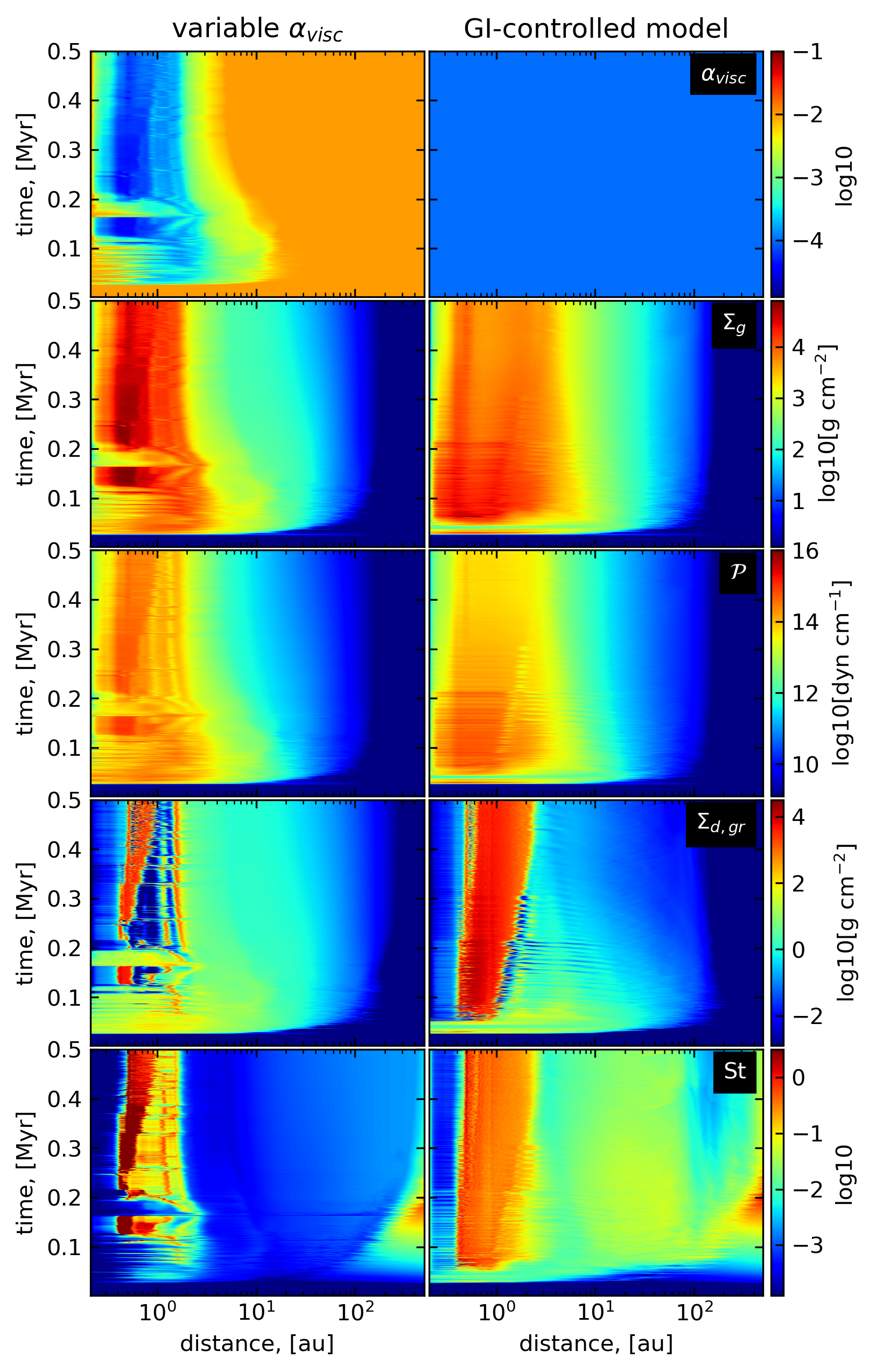} 
\par \end{centering}
\caption{Temporal evolution of the azimuthally-averaged disk characteristics in the model with variable $\alpha_{\rm visc}$ (left column) and the model with constant $\alpha_{\rm visc} = 10^{-4}$ (right column) from top to bottom: viscous $\alpha$-parameter, gas surface density, integrated pressure, grown dust surface density, and the Stokes number.} 
\label{fig:radial_alpvar}
\end{figure}

\subsection{Ring formation in the layered-disk model}
\label{sec:ring_layered}

A steady-state protoplanetary disk with a constant $\alpha$-parameter has a radial profile of $\Sigma_{\rm g}$ that monotonically increases toward the star. For typical conditions in a viscous disk, the scaling is $\Sigma_{\rm g} \propto r^{-1}$ \citep{Armitage2022}. This simple scaling may change if we consider a steady-state protoplanetary disk in the layered-disk model with a radially varying $\alpha$-parameter described by Eq.~(\ref{alpha:visc}).
The disk outer regions are usually characterized by the gas density that is low enough for the entire vertical column to be sufficiently ionized by cosmic rays for the MRI to operate. This makes the outer parts of the disk fully MRI-active with $\alpha_{\rm visc} \approx 10^{-2}$. As the gas density increases closer to the star, the MRI-dead regions may appear if the local column density of gas toward the disc midplane $\Sigma_{\rm g}/2$ exceeds the maximum thickness of the disk MRI-active layer $\Sigma_{\rm a}$. The thickness of the MRI-dead region further increases with increasing $\Sigma_{\rm g}$ (or decreasing distance $r$), which simultaneously lowers the effective $\alpha_{\rm visc}$ of the disk vertical column (see Eq.~\ref{alpha:visc}). 
Nevertheless, $\alpha_{\rm visc}$ retains a small but non-zero value in the  very dense regions due to the presence of residual viscosity $\alpha_{\rm rd}$, which is the result of hydrodynamic turbulence induced by Maxwell stress in the active disk layer \citep{Okuzumi2011}. 
Still closer to the star ($r<0.1$~au),
the disk temperature rises enough for the thermal ionization to set in ($T \geqslant 1300$~K), causing again the MRI activation in the entire vertical column and resulting in elevated values of $\alpha_{\rm visc}$ in the innermost parts of the disk. As shown in Appendix~\ref{app:steady_state}, the corresponding surface density profile becomes non-monotonic and features a gas density enhancement in the disk regions with lowest $\alpha_{\rm visc}$-values.

The main disk characteristics of the layered disk model are presented in the left column of Figure~\ref{fig:radial_alpvar}. 
The first panel shows the time evolution of the viscous  parameter $\alpha_{\rm visc}$, the behavior of which is  consistent with the analytical expectations. 
The disk outer parts are MRI-active with $\alpha_{\rm visc} = 10^{-2}$. The $\alpha_{\rm visc}$-parameter starts decreasing at $r < 10$~au, manifesting the formation of the dead zone. The deepest regions of the dead zone with $\alpha_{\rm visc} \leqslant 10^{-4}$ are located between 0.3 and 1.0~au. The radial extent of the dead zone in the early evolutionary stages is greater owing to the higher density of the disk.  
The early evolution is also characterized by notable horizontal spikes with high values of $\alpha_{\rm visc}\approx 0.01$ in the inner 2~au. These spikes are caused by the MRI bursts triggered by the thermal ionization of the dead zone. During these events matter accretes onto the star rapidly on a short viscous timescale, typically no more than a couple hundred years per event. The burst activity starts almost immediately after the disk formation and lasts up to $t \simeq 200$~kyr with a notable quiescent phase around 150~kyr.  The MRI-triggered bursts in the layered-disk model were considered in detail in \citet{Kadam2020}.

The radial gas surface density distribution is shown in the second panel of Figure~\ref{fig:radial_alpvar}.  The disk forms at about $t=0.027$~Myr after the onset of the gravitational contraction of the prestellar cloud when its spinning-up material hits the centrifugal barrier near the inner computational boundary. At this time instance, the gas surface density (but also $\Sigma_{\rm gr}$ and $\cal{P}$) features a sharp rise, reflecting the accumulation of matter in the disk, which quickly grows in size accompanied by fast dust growth.  After the disk formation instance,  $\Sigma_{\rm g}$ features a strong peak at the position of the dead zone, in agreement with the analytic expectations presented in Appendix~\ref{app:steady_state}. Mass and angular momentum are transported through the disk by the viscous torques at different rates, which are proportional to the radially varying values of $\alpha_{\rm visc}$. Fast transport in the outer disk with $\alpha_{\rm visc}=10^{-2}$ is followed by low transport in the inner disk where $\alpha_{\rm visc} \lesssim 10^{-3}$. As a result, a dead zone forms in which 
viscosity is not capable of carrying matter at a rate that matches that of the outer disk. Owing to this bottleneck effect the gas accumulates in the vicinity of the dead zone.
In the early stages of disk evolution multiple MRI bursts occur, which serve as an efficient mechanism of mass removal from the dead zone.   We note that at $t\lessapprox 0.1$~Myr the burst activity is so strong that the dead zone is frequently destroyed and reformed. After the end of the burst period, $t\geqslant 0.2$~Myr, gas shortly re-accumulates in the inner disk region  and the dead zone becomes stable afterwords.

The vertically integrated gas pressure is shown in the third panel of Figure~\ref{fig:radial_alpvar} and features a pressure maximum in the dead zone. The vertically integrated pressure is directly proportional to the product of the surface density and temperature,  and the formation of the pressure peak is not unexpected. We note, however, that in the dead zone $\alpha_{\rm visc}$ is low, which implies less viscous energy dissipation and hence lower temperatures, thus lowering the gas pressure as well. Nevertheless, the pressure bump does appear in the dead zone, although it is not as expressed as the surface density peak.

The fourth panel of Figure~\ref{fig:radial_alpvar} presents the surface density distribution of grown dust.
There are several local concentrations in the form of dense dust rings, the positions of which coincide with the local pressure maxima. The inner ring is located in the dead zone, while the outer one is at the outer edge of the gas accumulation region. It is known that grown dust concentrates in pressure bumps because of particle drift along the direction of increasing pressure  \citep[see e.g.][]{Weidenschilling1977, Armitage2001}.
The drift velocity is proportional to the pressure gradient and  the Stokes number $\mathrm{St}=t_{\rm stop} \,\Omega_{\rm K}$, where $t_{\rm stop}=\rho_{\rm s} a_{\rm max} / (\rho_{\rm g} c_{\rm s})$ is the stopping time, $\rho_{\rm g}$ the gas volume density, and $\Omega_{\rm K}$ the Keplerian velocity. The dust drift timescales become shorter than $10^{5}$ years for $\mathrm{St} \ge 10^{-2}$ \citep[see, e.g.,][]{Vorobyov2022}.
As the bottom panel in Figure~\ref{fig:radial_alpvar} demonstrates, the Stokes number approaches unity in the vicinity of the ring, which implies an efficient dust drift towards the local pressure maxima in the dead zone.





\subsection{Ring formation in the GI-controlled disk model}
\label{sec:ring_MRI_dead}

\begin{figure}
\begin{centering}
\includegraphics[width=\linewidth]{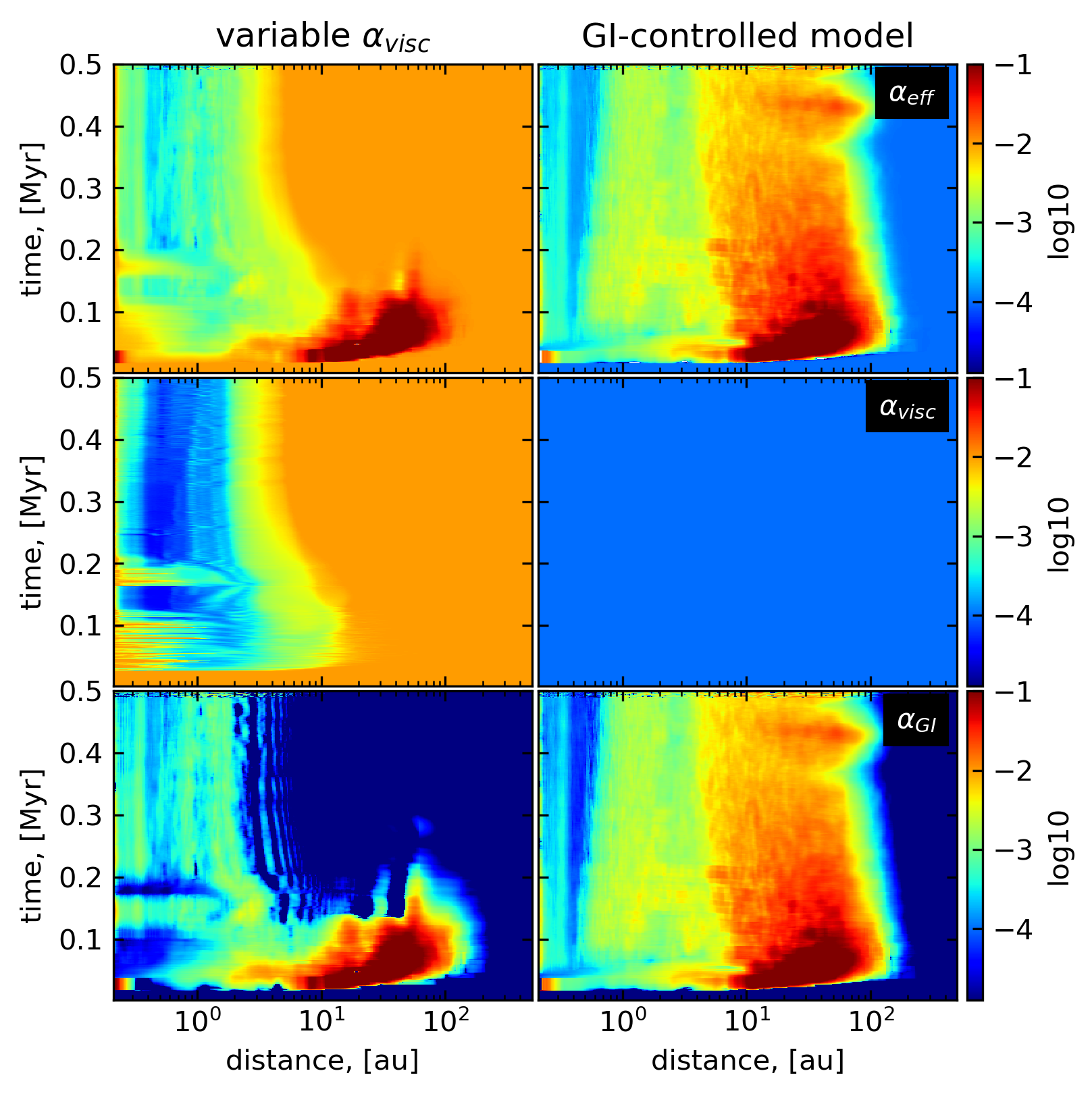} 
\par \end{centering}
\caption{Temporal evolution of the azimuthally averaged $\alpha$-parameters. The top, middle, and bottom panels show $\alpha_{\rm eff}$, $\alpha_{\rm visc}$, and $\alpha_{\rm GI}$, respectively. The left and right columns present the models with a radially variable $\alpha_{\rm visc}$ and spatially constant $\alpha_{\rm visc} = 10^{-4}$, respectively. } 
\label{fig:alpvisc_alpeff}
\end{figure}

\begin{figure*}
\begin{centering}
\includegraphics[width=\textwidth]{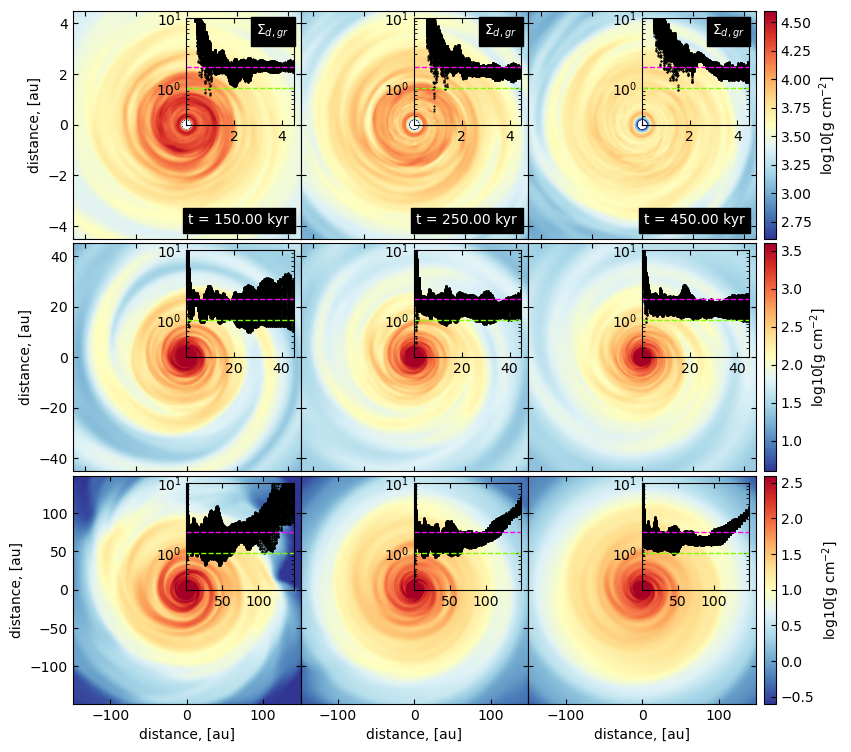} 
\par \end{centering}
\caption{Gas surface density in the GI-controlled model shown at different spatial scales and evolution times. The panels from top to bottom capture an increasingly larger spatial region, while the columns from left to right present show the disk of a progressively older age. The insets in each of the panels display the Toomre $Q$-parameter as a function of radial distance. The dashed pink and green lines correspond to $Q=2$ and $Q=1$ for convenience.} 
\label{fig:2D_GI}
\end{figure*}

\begin{figure}
\begin{centering}
\includegraphics[width=\linewidth]{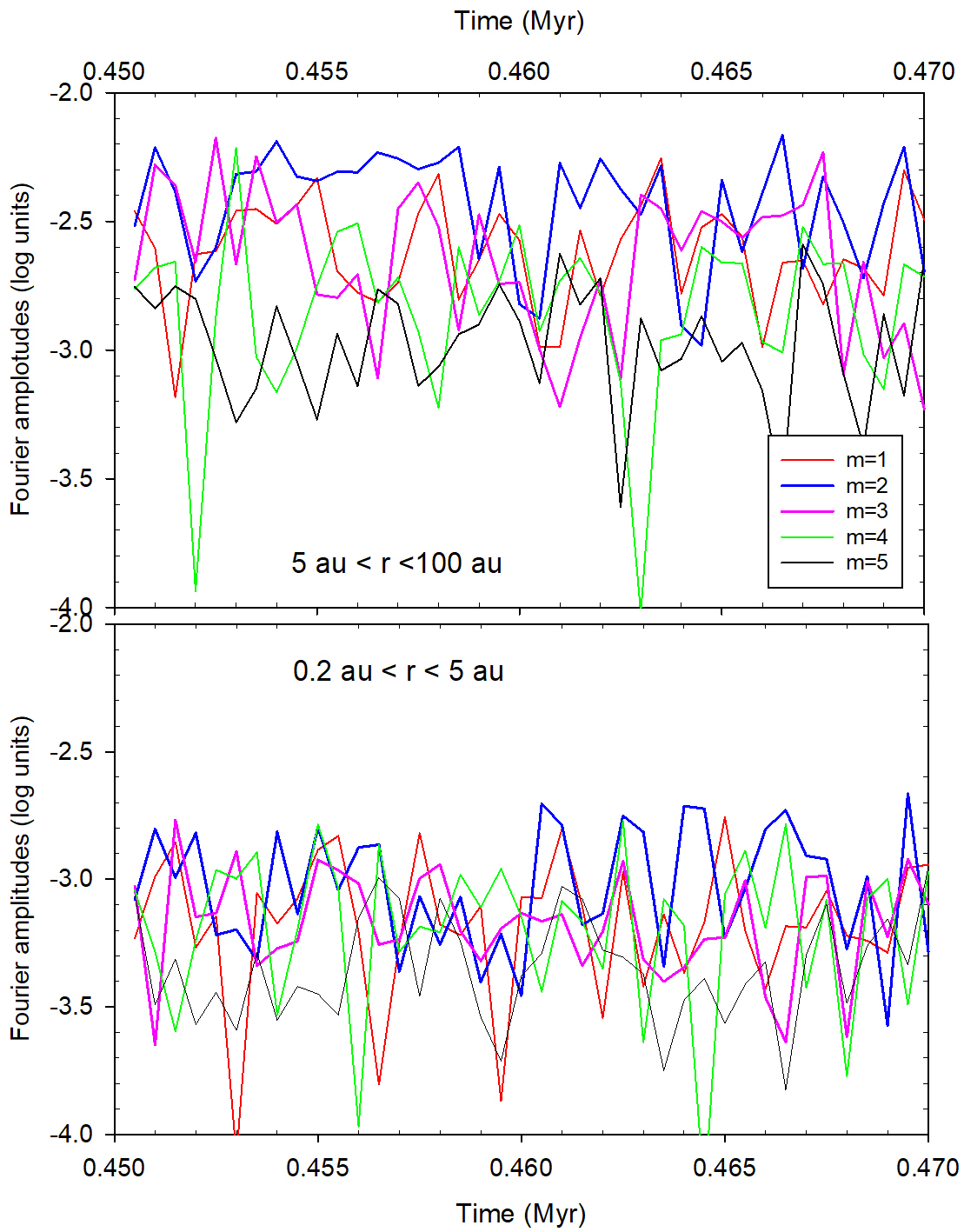} 
\par \end{centering}
\caption{Global Fourier amplitudes as a function of time. The top and bottom panels show the amplitudes for the outer and inner disk regions, respectively.}. 
\label{fig:fourier_modes}
\end{figure}

The dead zone development in the layered disk model is caused by a radially varying strength of the MRI in the disk, with high values of $\alpha_{\rm visc} \simeq 10^{-2}$ at $r\ge5$~au and low values ($ < 10^{-3}$) at $r\le 5$~au down to a fraction of astronomical unit, where the gas temperature is always high enough to sustain the MRI.
However, numerical studies suggest that the MRI may be suppressed by the nonideal MHD effects in almost the entire disk, except for its innermost parts \citep{BaiStone2013,Gressel2015}. Recent observations of efficient dust settling towards the disk midplane seem to support this theoretical finding  \citep{Zhang2018, Dullemond2018,Rosotti2020, DoiKataoka2021, Villenave2022}. In this case, the entire disk is formally a dead zone from the point of view of the layered disk model and it is not clear if dust can still accumulate in the inner disk regions.

To examine this case, we carried out the numerical simulation of a model disk with a suppressed MRI. We implemented this by setting the viscous $\alpha$-parameter to a small value $\alpha_{\rm visc} = 10^{-4}$ throughout the entire disk, implying that the MRI turbulence is significantly weakened as compared to the fully MRI-active case of $\alpha_{\rm visc}=10^{-2}$. 
The evolution of the GI-controlled model is presented in the right column of Figure~\ref{fig:radial_alpvar}. Interestingly, the model also demonstrates the accumulation of gas in the inner disk regions, although the accumulation zone is less sharp compared to the layered disk model. The pressure bump appears in the disk, the structure of which is smoother compared to the pressure bump in the layered disk model.  The single dust ring that forms in the GI-controlled disk just 15~kyr after the instance of disk formation is also notably wider than the corresponding rings in the layered disk. The Stokes number in the ring vicinity exceeds 0.1, which assists dust drift towards the local pressure maximum.

To understand the mechanism of the pressure bump and dust ring formation in the model with suppressed MRI, we note that the disk evolution in our models is governed not only by turbulent viscosity but also by disk self-gravity.  The latter can lead to the development of GI in sufficiently massive disks. The resulting gravitational torques may dominate the viscous torques in the early gravitationally unstable stages of disk evolution, especially when the $\alpha_{\rm visc}$ parameter is notably lower than $10^{-2}$ \citep{VorobyovBasu2009}.

To facilitate the comparison between the layered-disk model and the GI-controlled model,
we quantify the effect of gravity using the effective $\alpha_{\rm GI}$-parameter. 
First, we compute the gravitational stress in the disk plane as follows \citep{Riols2018}
\begin{equation}
\label{eq:grav_stress}
G_{\rm r \mathrm{\phi}} = \frac{1}{4 \pi G r} \frac{\partial \Phi}{\partial r} \frac{\partial \Phi}{\partial \phi}
\end{equation}
where $\Phi$ is the gravitational potential in the disk. We note that the non-zero stress is possible only if both the radial and azimuthal variations in $\Phi$ are present in the disk, which can be caused by gravitational instability or other global non-axisymmetric perturbations of the disk. The effective $\alpha_{\rm GI}$-parameter due to GI can then be expressed (by analogy to $\alpha_{\rm visc}$, see \citet{Kratter2016}) as
\begin{equation}
\alpha_{\rm GI} = {G_{\rm r\phi} \over P \left| \frac{d \ln \Omega_{\rm K}}{d \ln r} \right|},
\label{eq:alpha-GI}
\end{equation}
where $P$ is the gas pressure at the disk midplane (not to be confused with vertically integrated pressure $\cal P$ used in Eq.~(\ref{eq:mom}). To reduce the small-scale noise introduced by local variations in $G_{\rm r\phi}$ and $P$, we apply a running average to $\alpha_{\rm GI}$ at every grid cell with a time window of several thousand years.
Using $\alpha_{\rm GI}$ as a proxy for the efficiency of mass and angular momentum transport is justified for sufficiently massive disks with the disk-to-star mass ratio $\ge 0.2$ \citep{Vorobyov2010}, a condition satisfied by our model. Finally, we define the effective $\alpha_{\rm eff}$-parameter as the sum of the MRI and GI components
\begin{equation}
\alpha_{\rm eff}=\alpha_{\rm visc}+\alpha_{\rm GI}.
\end{equation}

The resulting radial distribution of $\alpha_{\rm eff}$ as a function of time is shown in the top row of Figure~\ref{fig:alpvisc_alpeff} for the layered and GI-controlled disk models. The middle and bottom rows show the corresponding distributions of $\alpha_{\rm visc}$ and $\alpha_{\rm GI}$ for comparison. We first consider the layered disk model shown in the left column of Figure~\ref{fig:alpvisc_alpeff}. The radial distributions of $\alpha_{\rm eff}$ and $\alpha_{\rm visc}$ in this model are qualitatively similar, though displaying some quantitative differences. Both $\alpha$-parameters are highest beyond 10~au and decline at smaller distances. This form of the $\alpha$-parameter distribution leads to the formation of a dead zone in the inner disk, as described in Sect.~\ref{sec:ring_layered}. The highest values of $\alpha_{\rm eff}\approx 10^{-1}$ between 10 and 100~au (red blob) are caused by a strong contribution from  $\alpha_{\rm GI}$ owing to strong gravitational instability in the early disk evolution.  We note, however, that the contribution quickly diminishes and already after 200~kyr the region beyond 10~au is dominated by turbulent viscosity due to MRI with $\alpha_{\rm eff}=10^{-2}$. This occurs because strong turbulent viscosity depletes and spreads out the disk, lowering $\Sigma_{\rm g}$ across the disk and reducing the strength of GI in the layered-disk model. However, GI does not disappear completely as evidenced by low but yet non-zero values of $\alpha_{\rm GI}$.   When the contribution from $\alpha_{\rm GI}$ to $\alpha_{\rm eff}$ is considered, the depth of the dead zone becomes shallower, but the contrast in the values of $\alpha_{\rm eff}$ between the dead zone and the rest of the disk is still considerable, exceeding a factor of 10.


We now consider the GI-controlled model with a suppressed MRI shown in the right column of Figure~\ref{fig:alpvisc_alpeff}. The radial distributions of $\alpha_{\rm visc}$ and $\alpha_{\rm eff}$ in the GI-controlled model are qualitatively different. While $\alpha_{\rm visc}$ is low and constant throughout the entire disk, $\alpha_{\rm eff}$ demonstrates strong radial variations. The highest values of $\alpha_{\rm eff}\sim 10^{-2}-10^{-1}$ are found  in the outer disk regions between 10~au and 100~au, and they notably decline in the inner disk to $\alpha_{\rm eff}\le 10^{-3}$. The overall form of the $\alpha_{\rm eff}$ parameter in the GI-controlled model suggests the formation of a dead zone in the inner disk, but the origin of the dead zone is now explained by the radial variations in $\alpha_{\rm GI}$, which has the dominant contribution to $\alpha_{\rm eff}$. We also note that the values of $\alpha_{\rm eff}$ in the GI-controlled model at 10--100~au gradually decline with time, reflecting a diminishing strength of GI with time, although it lasts longer than in the layered disk model.

To understand the origin of radial variations in $\alpha_{\rm GI}$ (and hence in $\alpha_{\rm eff}$) in the GI-controlled model, we show in Figure~\ref{fig:2D_GI} the corresponding gas surface density distribution at different spatial scales.  The disk remains gravitationally unstable and exhibits a developed spiral structure throughout the entire evolution period covered by our simulation, although the sharpness of the spiral pattern weakens with time.  To describe the propensity of a disk to develop gravitational instability, the Toomre parameter is usually used. When the dust component is present in the gas disk, the Toomre parameter can be defined as follows \citep{2018VorobyovAkimkin}
\begin{equation}
    \label{eq:ToomreQ}
    Q = \dfrac{\tilde{c}_{\rm s} \Omega_{\rm g}}{\pi G \left( \Sigma_{\rm g} + \Sigma_{\rm d, tot} \right)},
\end{equation}
where $\tilde{c}_{\rm s} = c_{\rm s} \sqrt{1 + \xi_{\rm d2g}}$ is the modified sound speed and $\Sigma_{\rm d,tot} = \Sigma_{\rm d.gr} + \Sigma_{\rm d,sm}$ the total surface density of dust.


The insets in Figure~\ref{fig:2D_GI} show the radial distributions of the $Q$-values for all grid zones at a given radius with the corresponding spatial scale preserved. The characteristic values below which the disk tends to develop gravitational instability ($Q \lesssim 2$) and fragmentation ($Q \lesssim 1$) are shown by the pink and green horizontal dashed lines, respectively. Clearly, the disk satisfies the Toomre $Q \lesssim 2$ criterion throughout the considered evolution period. A decrease in the gas density owing to accretion onto the central star in the course of evolution is compensated by a matching decrease in the disk temperature owing to the lowering optical depth of the disk. 

We note, however, that the $Q$-parameter sharply increases in the innermost disk regions ($r \le 1.0-2.0$~au) and also in the regions beyond the disk extent ($r>100$~au). The latter is caused by a sharp drop in the gas surface density beyond the disk outer edge, while the former is caused by strongly increasing sheer (as represented by $\Omega_{\rm g}$) and gas temperature (as represented by $\tilde{c}_{\rm s}$) in the inner disk. This behaviour of the $Q$-parameter was also seen in other numerical hydrodynamics simulations of purely gaseous disks \citep{Bae2014}. The sharp rise of the $Q$-parameter at $r< 1.0-2.0$~au and the corresponding weakening of gravitational instability can explain the decrease in $\alpha_{\rm GI}$ seen in Figure~\ref{fig:alpvisc_alpeff} in the inner disk. 

We can quantify the effect of a radially varying strength of gravitational instability in terms of the global Fourier amplitudes defined as
\begin{eqnarray}
C^{\rm in}_{\rm m} (t) &=& {1 \over M_{\rm d}} \left| \int_0^{2 \pi} 
\int_{r_{\rm sc}}^{5\mathrm{au}} 
\Sigma_{\rm g}(r,\phi,t) \, e^{im\phi} r \, dr\,  d\phi \right|, \\
C^{\rm out}_{\rm m} (t) &=& {1 \over M_{\rm d}} \left| \int_0^{2 \pi} 
\int_{r_{5\mathrm{au}}}^{100\mathrm{au}} 
\Sigma_{\rm g}(r,\phi,t) \, e^{im\phi} r \, dr\,  d\phi \right|,
\label{fourier}
\end{eqnarray}
where $M_{\rm d}$ is the disk mass and $m$ is the spiral mode.  
The Fourier amplitudes can be regarded as a measure of the perturbation amplitude of spiral density waves in the disk compared to the underlying axisymmetric density distribution.
When the disk surface density is axisymmetric, the amplitudes of all modes are equal to zero. With this definition, $C^{\rm in}_{\rm m}$ and $C^{\rm in}_{\rm m}$ represent the Fourier amplitudes of the inner (0.2--5.0~au) and outer (5.0--100~au) disk regions. This spatial division roughly traces a sharp change in the $\alpha_{\rm eff}$-values as seen in the GI-controlled model (upper right panel in Fig.~\ref{fig:alpvisc_alpeff}). 

Figure~\ref{fig:fourier_modes} presents the Fourier amplitudes $C^{\rm in}_{\rm m}$ and $C^{\rm in}_{\rm m}$ calculated during a time interval of 20~kyr. The Fourier amplitudes confirm that the gravitational instability is stronger in the disk region between 5.0 and 100~au as compared to the disk interior to 5.0~au. The dominant $m=2$ mode in the outer disk is almost an order of magnitude higher than the strongest mode in the inner disk. The behavior of Fourier amplitudes at other evolutionary times is similar.

\section{Parameter space study}
\label{paramspace}

\begin{figure}
\begin{centering}
\includegraphics[width=\linewidth]{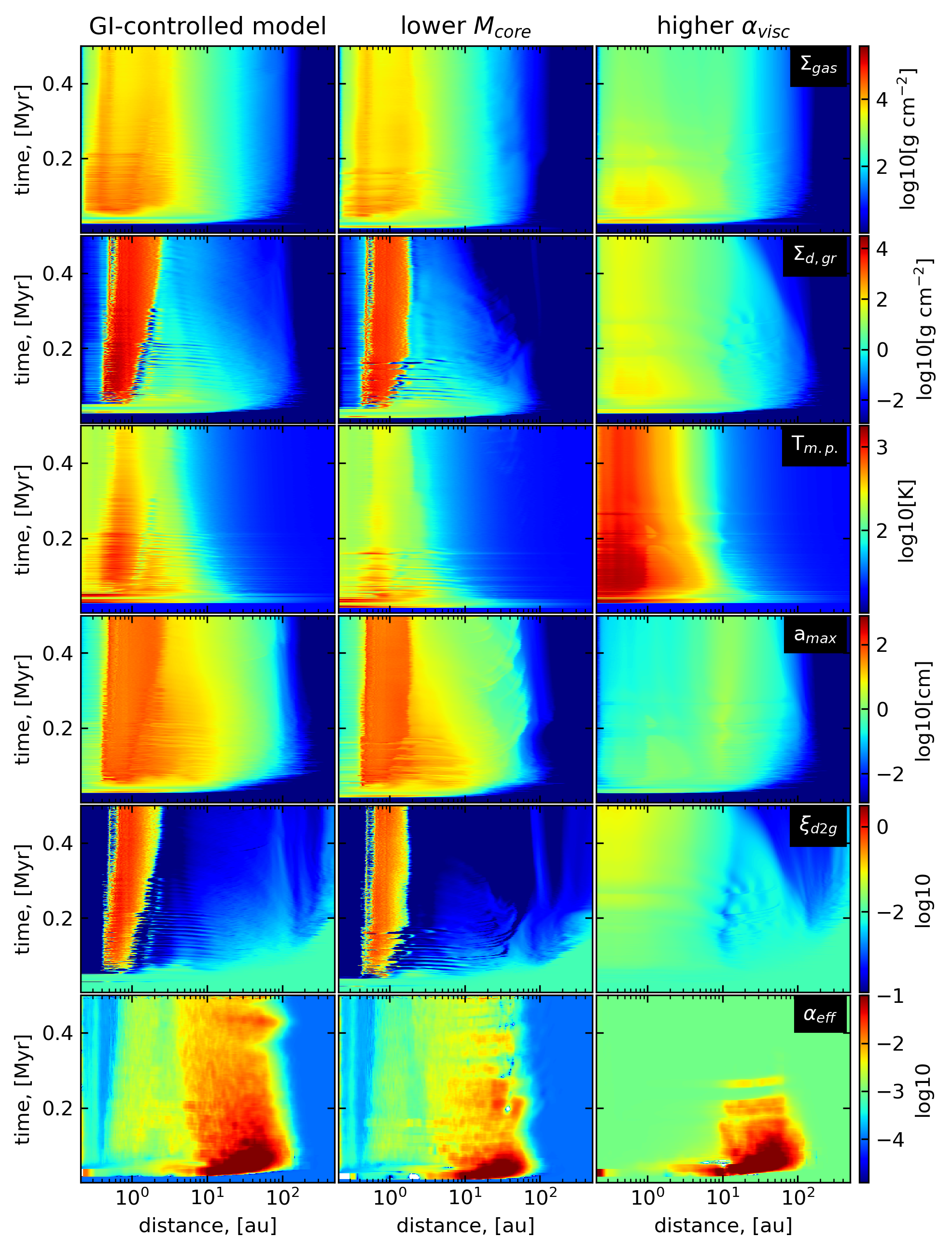} 
\par \end{centering}
\caption{Space-time plots showing the time-evolution of the azimuthally averaged disk characteristics. The columns from left to right correspond to the fiducial model with $M_{\rm core}=0.53~M_\odot$ and $\alpha_{\rm visc}=10^{-4}$, the model with a lower $M_{\rm core}=0.3~M_\odot$, and the model with higher $\alpha_{\rm visc}=10^{-3}$. The rows from top to bottom show: the gas surface density, grown dust surface density, gas temperature in the disk midplane, maximum dust size, total dust-to-gas mass ratio, and the effective $\alpha$-parameter. } 
\label{fig:param-GI}
\end{figure}

Here, we consider the effects of variations in the initial cloud core mass and $\alpha_{\rm visc}$ on the efficiency of dust trapping in the GI-induced ring. Figure~\ref{fig:param-GI} presents the azimuthally averaged disk characteristics as a function of time for our fiducial model and two more models: one with almost a factor of two smaller initial cloud core mass ($M_{\rm core}=0.3~M_\odot$) and the other with a larger MRI turbulence as represented by a spatially constant value of $\alpha_{\rm visc}=10^{-3}$. The former is to probe if lower mass cores can still form disks that are capable of supporting GI and forming GI-induced dust rings. The latter is to demonstrate the critical effect of the MRI turbulence in the suppression of GI-induced rings. The second column in Figure~\ref{fig:param-GI} demonstrates that prestellar cores with mass as low as $0.3~M_{\odot}$ can still form disks that sustain GI and lead to the formation of GI-induced dust rings around 1~au. The dust ring is somewhat narrower and of lower density, which results in lower temperatures in the ring vicinity owing to lower optical depths. 
The bottom row in Figure~\ref{fig:param-GI} displays the $\alpha_{\rm eff}$-parameter as the sum of $\alpha_{\rm visc}$ and $\alpha_{\rm GI}$. The strongest positive radial gradient in $\alpha_{\rm eff}$ across the gas disk extent is found for the fiducial model. This model is also characterized by the strongest dust ring. The model with a lower $M_{\rm core}$ has a weaker gradient of $\alpha_{\rm eff}$ (especially at later evolution times), owing to a weaker GI in a less massive disk. This results in a ring with smaller dust-to-gas mass ratios compared to the fiducial model. Although we have compared only two simulations with different initial cloud core masses, these two simulations lead to similar results and suggest that in this range of initial core masses dust trapping remains similar.

\begin{figure*}
\begin{centering}
\includegraphics[width=\linewidth]{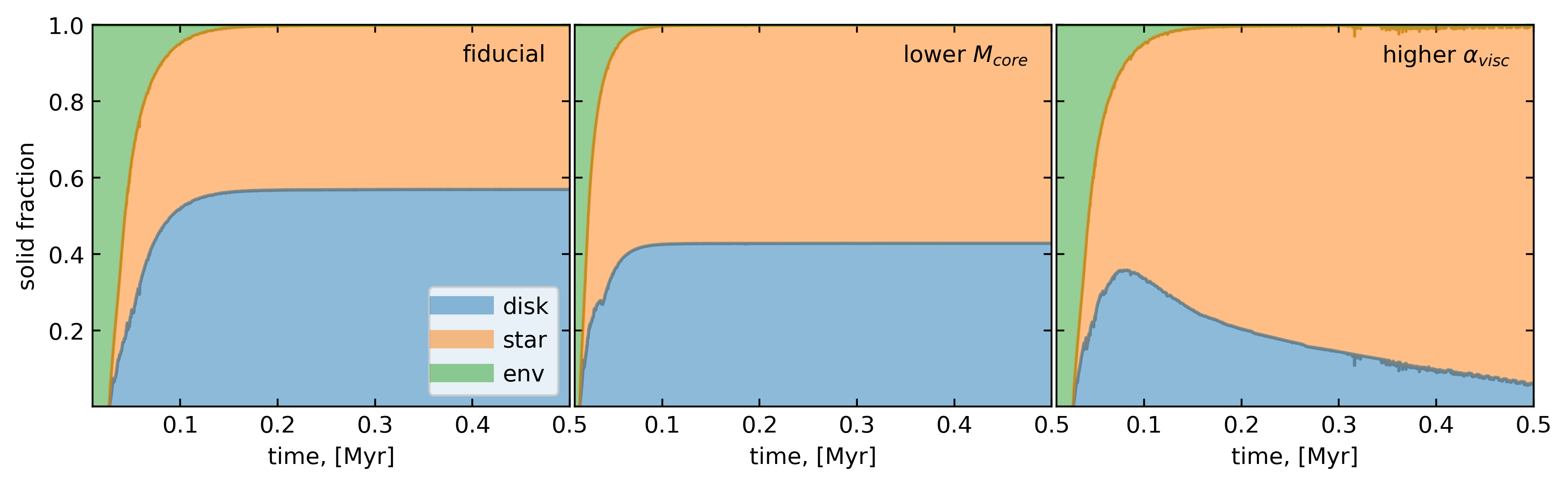} 
\par \end{centering}
\caption{Fractions of the total dust mass budget contained in the disk (blue), envelope (green), and drifted to the star (orange). The panels from left to right correspond to the  fiducial model with $M_{\rm core}=0.53~M_\odot$ and $\alpha_{\rm visc}=10^{-4}$, the model with a lower $M_{\rm core}=0.3~M_\odot$, and the model with higher $\alpha_{\rm visc}=10^{-3}$.} 
\label{fig:tot-mass-budget}
\end{figure*}

The picture qualitatively changes  when the model with a higher value of $\alpha_{\rm visc}$ is considered. In this case, the sharp dust ring around 1~au is replaced with a abroad dust density enhancement in the inner several au. The values of $\xi_{\rm d2g}$ can be as high as 0.09, but they are still much lower than the corresponding values in the other two models with $\alpha_{\rm visc}=10^{-4}$. The disk temperature in the inner several au rises notably because of more efficient viscous heating in the disk midplane. This qualitative change in the dust dynamics can be understood from the radial distribution of the effective $\alpha$-parameter. The model with higher $\alpha_{\rm visc}$ has no clear radial gradient in $\alpha_{\rm eff}$. Instead, it has a strong enhancement in $\alpha_{\rm eff}$, which is localized in time and space to the initial 0.2~Myr of disk evolution and to a radial annulus $r\simeq 10-100$~au. This means that the input of GI to the mass and angular momentum transport is limited to the intermediate and outer disk regions and to the initial stages of disk evolution.  The rest of the disk extent and the evolution time is controlled by turbulent viscosity due to MRI, which is assumed to be constant in time and space. Such a disk features no compact dead zones. For larger values of $\alpha_{\rm visc}$, the effect is even stronger and the dust accumulation mostly vanishes.


The effect of varying $\alpha_{\rm visc}$ can be understood as follows. The dust drift velocity in the disk is composed of two components: the gradiental drift that depends on the local pressure gradient and the advective drift that depends on the value of $\alpha$-parameter \citep{2016Birnstiel}.   
As was shown in \citet{Vorobyov2023a}, an increase in $\alpha_{\rm visc}$ acts to increase the advective drift velocity, which is generally pointed towards the star in GI-unstable disks, while the gradiental drift velocity is weakly affected. The dust particles are now less efficiently trapped by the local pressure bumps, and more dust now drifts across the inner disk and onto the star. The net result is the reduction in the dust accumulation efficiency in the disk. For $\alpha_{\rm visc}=10^{-2}$, dust drift is dominated by advection with the gas flow \citep{Vorobyov2023a}.

Our interpretation is confirmed with the analysis of the dust mass budget in the system shown in Figure~\ref{fig:tot-mass-budget}. In particular, the fractions of the dust mass contained in the disk, envelope, and also drifted through the inner sink cell are plotted as a function of time in the models considered. We do not follow the fate of the latter component, simply assuming that this fraction is sublimated and the resulting refractory species land on the star. Clearly, the fiducial model is most efficient in retaining dust in the disk, while the model with higher $\alpha_{\rm visc}$ loses most of its initial dust budget to the star. This trend is in agreement with our preceding analysis and with the strength of the dust rings found in the models.

\section{Prospects for the streaming instability}
\label{Sect:stream-inst}

\begin{figure}
\begin{centering}
\includegraphics[width=\linewidth]{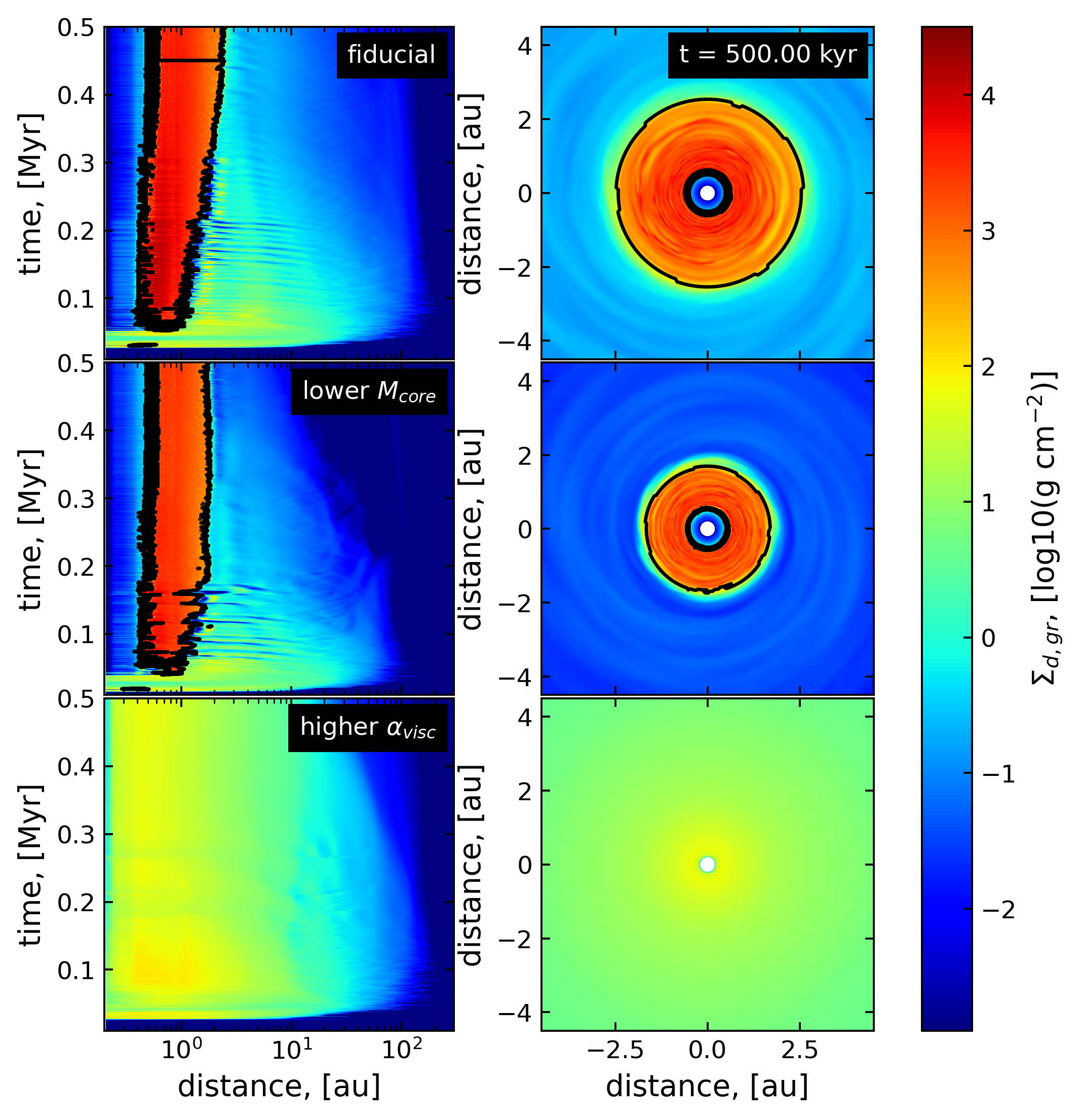} 
\par \end{centering}
\caption{Surface density of grown dust with the disk regions susceptible to the streaming instability identified by the black curves. The rows from top to bottom correspond to the fiducial model, model with a lower mass of the prestellar core, and model with a higher value of the viscous parameter.  The left column shows the time-dependent evolution of the azimuthally averaged $\Sigma_{\rm d,gr}$, while the right column presents the two-dimensional distribution of $\Sigma_{\rm d,gr}$ in the inner $10\times10$~au box at the end of simulations.} 
\label{fig:SI_evo}
\end{figure}

Dust rings such as those formed in the layered disk and GI-controlled models may be favorable sites for planetesimal formation via the process known as the streaming instability \citep[e.g.,][]{Youdin2005,Yang2017,Carrera2022}. Since the dust ring  in the GI-controlled model forms as early as 15~kyr after the disk formation instance, the resulting generation of planetesimals may represent the first building blocks of planets. Direct modeling of the streaming instability is difficult in the current work, since it requires a higher spatial resolution, and also dust and gas dynamics in the vertical direction (neglected in our thin-disk models). However, we can take the criteria obtained with proper high resolution modeling and apply them to our model disk to find out if it can be prone to develop the streaming instability. In particular, we take the following criteria presented in \citet{Yang2017}:
\begin{eqnarray}
\label{eq:SI_cond_1}
\log{\xi_{\rm d2g}} \geqslant 0.10 \left(\log{\mathrm{St}}\right)^2 + 0.20 \log{\mathrm{St}} - 1.76    \ \ \ \  (\mathrm{St} < 0.1), \\
\log{\xi_{\rm d2g}} \geqslant 0.30 \left( \log{\mathrm{St}}\right )^2 + 0.59 \log{\mathrm{St}} - 1.57   \ \ \ \  (\mathrm{St} > 0.1).
\label{eq:SI_cond_2}
\end{eqnarray}
These conditions are complemented by the requirement that the volume density of grown dust in the disk midplane $\rho_{\rm d.gr.}$ be equal to or greater than that of gas $\rho_{\rm g}$ \citep{Youdin2005}
\begin{equation}
    \zeta={\rho_{d.gr}\over \rho_{\rm g}} \ge 1.0.
    \label{Eq:plt_vol_dens}
\end{equation}
Here, the volume densities of grown dust and gas are calculated using the corresponding local vertical scale heights $H_{\rm d}$ and $H_{\rm g}$. This condition requires efficient dust settling in the disk. Although dust settling is not directly modeled with \simname{FEOSAD}, we can predict its efficiency from the known model parameters using Equation~(\ref{eq:dust-scale-height}) and assuming a Gaussian distribution of gas and dust in the vertical direction.
Depending on the local conditions in the disk, these criteria may or may not be fulfilled. 

In Figure~\ref{fig:SI_evo} we present the time evolution of the azimuthally averaged surface density of grown dust in the three considered models and also the spatial distribution of $\Sigma_{\rm d,gr}$ of the inner disk regions comprising the dust ring, taken at the end of simulations. The black curves delineate the disk zones in which the conditions for the development of the streaming instability are satisfied. Clearly, the dust rings in the fiducial and lower $M_{\rm core}$ models are prone to  develop the streaming instability starting from the ring formation instance and during the entire considered evolution period. However, the model with higher $\alpha_{\rm visc}$ fails to fulfil the streaming instability criteria, namely, the condition on the efficient dust settling (Eq.~\ref{Eq:plt_vol_dens}). An increase in $\alpha_{\rm visc}$ to $10^{-3}$ implies a reduced efficiency of dust settling, which impedes the development of the streaming instability under our assumptions. In a follow-up paper we will study the consequences of the streaming instability on the dust ring appearance and estimate the efficiency of planetesimal formation in the GI-controlled dust rings. 

\begin{figure}
\begin{centering}
\includegraphics[width=\linewidth]{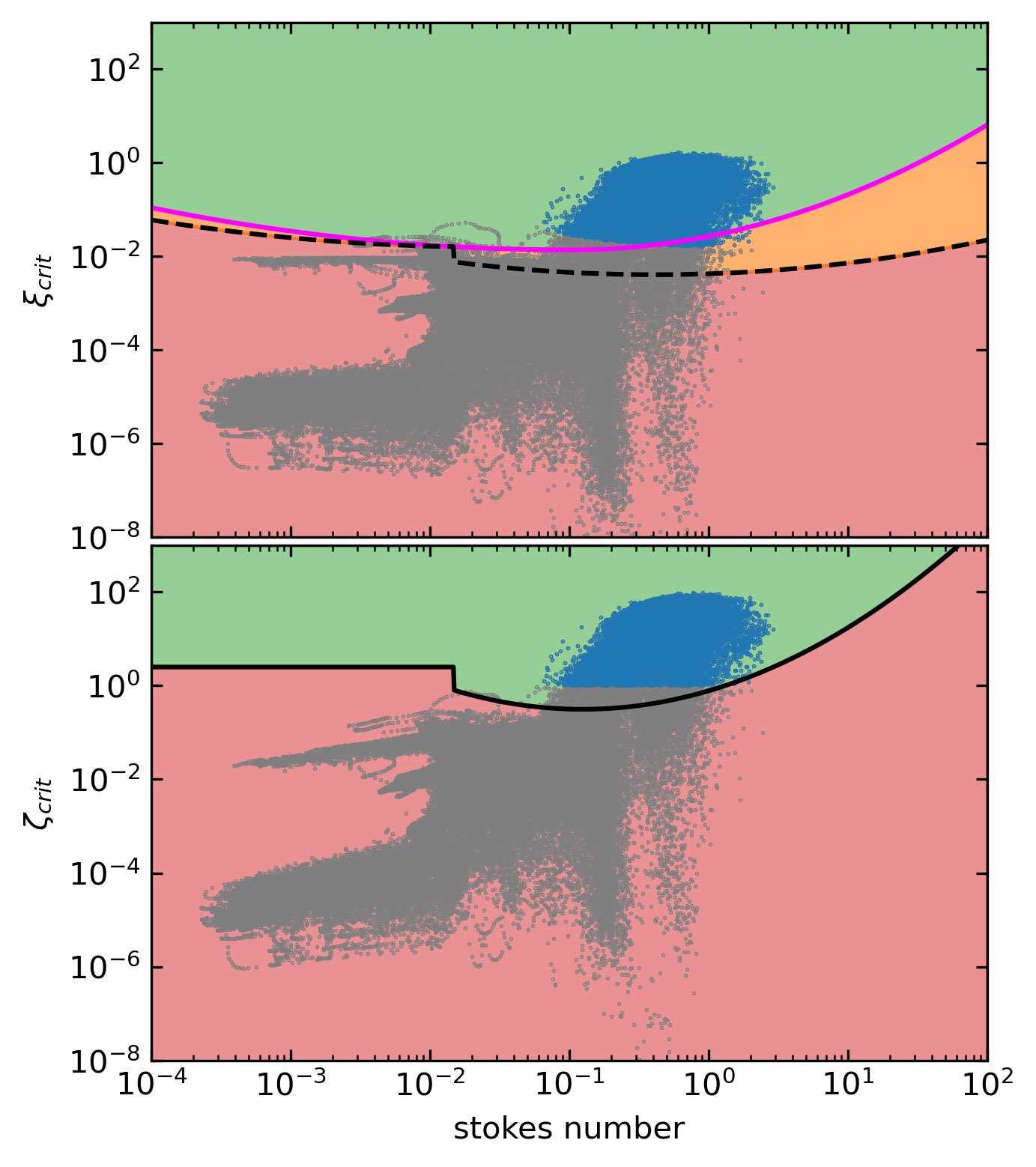} 
\par \end{centering}
\caption{ Streaming instability phase space. The top panel shows the ratio of the surface densities of grown dust to gas as a function of the Stokes number. The pink and dashed black lines depict the critical values for the development of the streaming instability according to \citet{Yang2017} and \citet{LiYoudin2021}. The data of the fiducial model are overlaid with filled circles, with blue ones fulfilling in addition the criterion on the ratio of volume densities (see Eq.\ref{Eq:plt_vol_dens}). The bottom panel shows panel displaces the ratio of the volume densities of grown dust to gas as a function of $\mathrm{St}$. The solid black line indicates the critical values for the streaming instability according to Eq.~(\ref{eq:Li2021-zeta}). The fiducial model data are overlaid with the grey circles.} 
\label{fig:phase-space-SI}
\end{figure}

{
To verify that the conditions for the streaming instability are fulfilled in our fiducial model, we plot in the top panel of Figure~\ref{fig:phase-space-SI} the critical values of $\xi_{\rm d2g}$ as a function of $\mathrm{St}$ according to \citet{Yang2017}, as laid out by Equations~(\ref{eq:SI_cond_1}) and (\ref{eq:SI_cond_2}). The corresponding values are shown with the pink curve, with the region above this cure being prone to develop the streaming instability. 

In addition, we also consider the more recent criterion for the streaming instability put forward in \citet{LiYoudin2021}
\begin{equation}
\label{eq:Li2021-xi}
    \log \left( {\xi_{\rm d2g} \over \Pi} \right) =  A \left( \log \mathrm{St} \right)^2 + B \log \mathrm{St} + C
\end{equation}
where 
\begin{eqnarray}
    & &A=0.1, \, B=0.32, \, C=-0.24 \,\,\, \mathrm{if} \, \, \mathrm{St} < 0.015, \nonumber \\
    & &A=0.13, \, B=0.1, \, C=-1.07 \,\,\, \mathrm{if} \, \,\mathrm{St}>0.015. \nonumber
\end{eqnarray}
Here, $\Pi=0.05$ is the radial pressure gradient. We note that the value of $\Pi$ may vary in the disk, but we take it equal to 0.05 for our model data for consistency with the work of \citet{LiYoudin2021}.
The corresponding critical values for the streaming instability are plotted with the black dashed curve.  The condition on the streaming instability provided by \citet{LiYoudin2021} is milder than that of \citet{Yang2017}.

The data of the fiducial model are overlaid on the top panel of Figure~\ref{fig:phase-space-SI}, with each filled circle corresponding to the azimuthally averaged $\xi_{\rm d2g}$ and $\mathrm{St}$ for radial annuli  of our numerical grid that are located inside 150~au (the approximate disk extent). The entire disk evolution is considered with a time sampling of 500~yr. The difference between the grey and blue circles is that the latter also fulfill the condition on the ratio of volume densities in the disk midplane, as laid out by Equation~(\ref{Eq:plt_vol_dens}). 
As the top panel in Figure~\ref{fig:phase-space-SI} indicates, a certain fraction of the model data fulfils the imposed criteria and the streaming instability can indeed develop in our model disk

 Furthermore, we consider the updated criterion also provided in \citet{LiYoudin2021} but formulated in terms of the ratio $\zeta_{\rm crit}$ of the dust and gas volume densities in the disk midplane 
\begin{equation}
\log \zeta_{\rm crit} = A^\prime \left(\log \mathrm{St} \right)^2 + B^\prime \log \mathrm{St} + C^\prime,
\label{eq:Li2021-zeta}
\end{equation}
with
\begin{eqnarray}
    & &A^\prime=0, \, B^\prime=0, \, C^\prime=2.5 \,\,\, \mathrm{if} \, \, \mathrm{St} < 0.015, \nonumber \\
    & &A^\prime=0.48, \, B^\prime=0.87, \, C^\prime=-0.11 \,\,\, \mathrm{if} \, \,\mathrm{St}>0.015. \nonumber
\end{eqnarray}
The corresponding values in the $\zeta$ vs. $\mathrm{St}$ phase space are plotted in the bottom panel of Figure~\ref{fig:phase-space-SI} with the black solid line showing the critical values for the development of the streaming instability. This new criterion is also fulfilled in our fiducial model.

To better quantify the feasibility of planetesimal formation in the fiducial model, we calculated the dust mass in the disk that is prone to the development of the streaming instability, $M_{\rm d,gr}(\mathrm{SI})$. In addition, we also calculated the aria of the disk that encompasses the disk regions prone to develop the streaming instability, $\mathrm{Area(SI)}$. Each value is normalized either to the total mass of grown dust or to the disk area, assuming, for simplicity, that the disk radius is 150~au (see Sect~\ref{Sect:implications}).
While the disk area within which the streaming instability can operate is only a minor fraction of the total area occupied by the disk, the corresponding dust mass that is prone to the streaming instability is a large fraction of the total dust mass in the disk, reflecting efficient dust drift and accumulation in the GI-induced dead zone.
}

\begin{table}[]
    \centering
    \begin{tabular}{|c|c|c|}
        \hline
        Streaming instability condition & $\mathrm{Area(SI)}$ & $M_{\rm d,gr}(\mathrm{SI})$ \\ \hline \hline
         \citet{Yang2017}, eqs.~(\ref{eq:SI_cond_1}), (\ref{eq:SI_cond_2}), (\ref{Eq:plt_vol_dens}) & 0.0268\% & 87.9\% \\ \hline
         \citet{LiYoudin2021}, eqs.~(\ref{eq:Li2021-xi}), (\ref{Eq:plt_vol_dens}) & 0.0269\% & 87.91\% \\ \hline
         \citet{LiYoudin2021}, eq.~(\ref{eq:Li2021-zeta}) & 0.0288\% & 88.21\% \\ \hline 
    \end{tabular}
    \caption{ Middle column is the total disk area where the conditions for the streaming instability are satisfied relative to the total area of the disk. Right column -- the total mass of grown dust contained within the disk region prone to the streaming instability relative to the total mass of grown dust in the disk. All data are for the fiducial model.}
    \label{tab:SI_percentage}
\end{table}

\section{Implications for dust disk sizes and masses}
\label{Sect:implications}

\begin{figure}
\begin{centering}
\includegraphics[width=\linewidth]{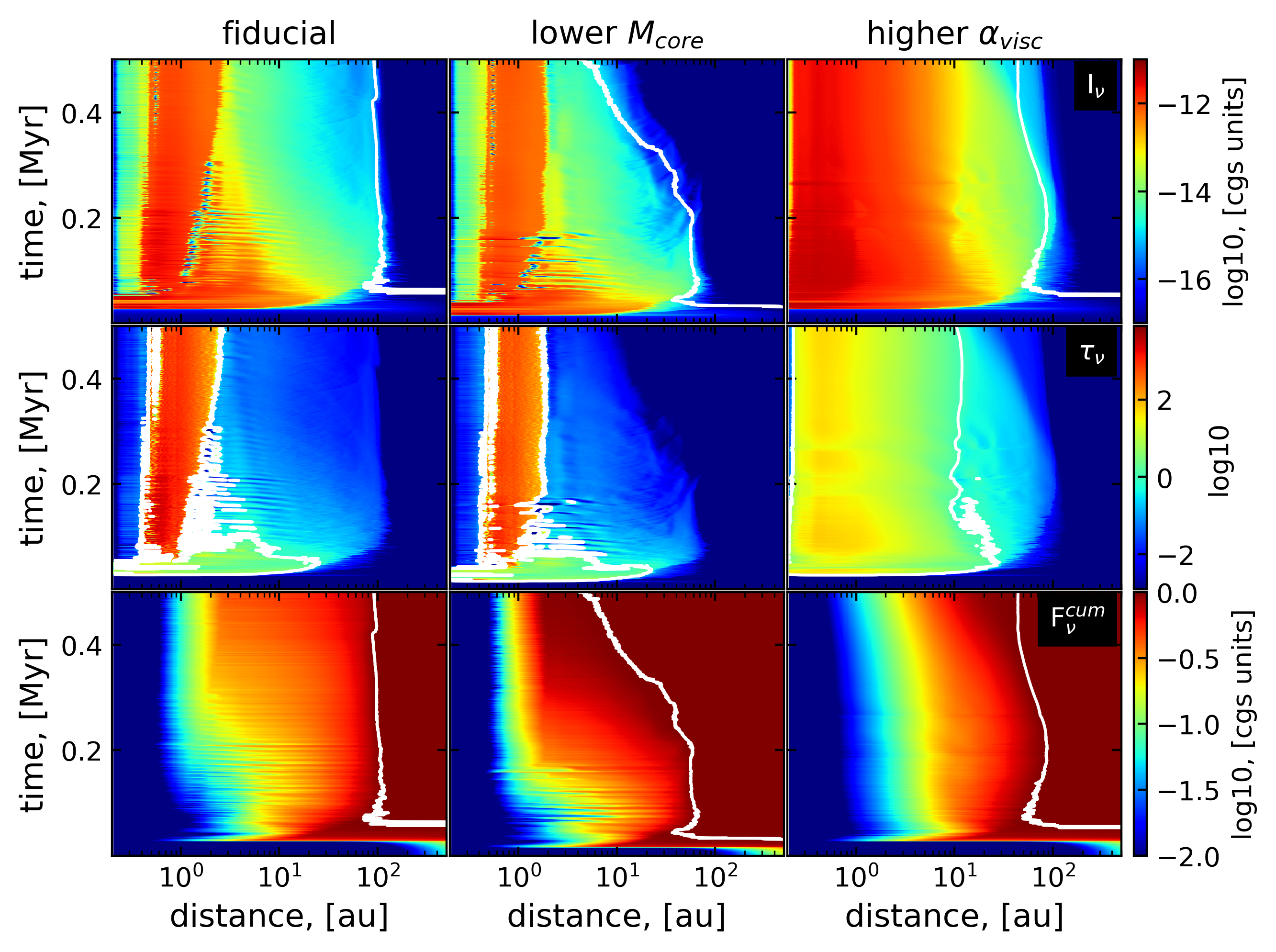} 
\par \end{centering}
\caption{Space-time plots showing the evolution of dust radiation intensity (top row), optical depth (middle row), and cumulative flux (bottom panel) at 3~mm. Columns from left to right correspond to the fiducial model, model with a lower $M_{\rm core}$, and model with higher $\alpha_{\rm visc}$. The white contours in the top and bottom rows delineate the radial locations, within which 95\% of the total flux is contained. The white curves in the middple row highlight the regions with the optical depth $> 1.0$. The units for radiation intensity and flux are erg~cm$^{-2}$~s$^{-1}$ Hz$^{-1}$ sr$^{-1}$  and erg~cm$^{-2}$~s$^{-1}$ Hz$^{-1}$, respectively.} 
\label{fig:mock-fluxes}
\end{figure}

Disk masses and radii play a key role in many physical processes responsible for mass and angular momentum transport, dust drift and growth, and planet formation. Yet, their observational estimates are associated with uncertainties, which may significantly alter the true disk masses and radii and lead to wrong conclusions \citep[e.g.,][]{2014DunhamVorobyov}.
We demonstrate this using our model disk as an example. The distribution and properties of dust in the fiducial model are known from simulations and we use them to calculate the underlying disk mass and size. We compare these ``true'' values with those derived using the methods and techniques applied when analysing the observations of real protoplanetary disks as described below.

We adopt a simplified model to calculate the radial distribution of the dust radiation intensity assuming a local plane-parallel disk geometry and dust temperature that is constant (or weakly changing) in the vertical direction.
We note that in our model we make no distinction between the gas and dust temperatures, which is justified for the bulk of the disk midplane at the solar metallicity \citep{Vorobyov2020c}, where most of the dust mass is supposed to reside due to vertical settling. We also note that in the plane of the disk, the temperature was computed self-consistently  using the vertically integrated gas pressure and gas density in each computational cell as $T_{\rm mp}=\mu {\cal P}/(\Sigma_{\rm g} {\cal R}$), where $\mu=2.33$ is the mean molecular weight and $\cal R$ is the universal gas constant. A formal solution of the radiative transfer equation in the plane-parallel limit can be written as
\begin{equation}
 I_{\nu}(r,\phi)=B_{\nu}(T_{\rm mp})(1-e^{-\tau_{\nu}}),
 \label{eq:modified-black-body}
\end{equation}
where $I_{\nu}$ is the radiation intensity at a given position ($r,\phi$) in the disk, $B_{\nu}(T_{\rm mp})$ is the Planck function, and 
$\tau_{\nu}=\kappa_{\rm \nu} (\Sigma_{\rm d, sm}+\Sigma_{\rm d, gr})$
is the total optical depth of the small and grown dust populations. The frequency dependent absorption opacity $\kappa_{\rm \nu}$  (per gram of dust mass)  for the small and grown dust populations with size range  from $5\times 10^{-3}$~$\mu$m to $a_{\rm max}$, respectively, were found using the \simname{OpacityTool} of \citet{Woitke2016} based on the Mie theory assuming pure silicate grains of spherical shape. The spatially resolved fluxes $F_{\nu}(r,\phi)$ for the assumed distance $d$ to the source, together with $I_{\rm \nu}(r,\phi)$, represent our mock observations. For a particular wavelength, we choose 3~mm, which corresponds to Band 3 on ALMA.

Figure~\ref{fig:mock-fluxes} presents the synthetic intensities and optical depths at 3~mm for the three models considered. In addition, the bottom panel displays the cumulative flux in the radial direction as a fraction of the entire flux contained within 500~au.  
We note that the logarithmic scale in the radial direction distorts the view and exaggerates the inner regions, which are hard to resolve otherwise.
The distance is set equal to $d=500$~pc.

The dust ring in the fiducial and lower $M_{\rm core}$ models is characterized by high optical depths and the corresponding intensity of radiation is dominated by the Planck function. On both sides of the ring, the disk becomes optically thin, so that $I_{\nu}$ also drops substantially. At $t \le 0.2$~Myr in the fiducial model and at $t\le 0.1$~Myr for the lower $M_{\rm core}$, the flux coming from the dust ring contributes only about 10\% to the cumulative flux owing to the small surface area of the ring compared to the rest of the disk. Most of the flux is coming from the disk regions outside the dust ring at this evolutionary stage. At later stages, as more dust drifts from the disk towards the inner ring, the contribution of the latter to the total flux increases to 25-30\%. Only after $t=0.5$~Myr the dust ring in the model with low $M_{\rm core}$ begins to dominate the cumulative flux.  
The higher $\alpha_{\rm visc}$ model is also characterized by optically thick inner regions up to about 10~au. However, the spatial distribution of $I_\nu$ is much smoother than in the other two models. The inner several astronomical units also provide a minor contribution to the total flux (about 10\%), which is dominated by the intermediate and outer disk regions. We also note that disks in all models feature a sharp outer edge in the spatial distribution of $I_{\nu}$.

We further calculate the dust disk radii and masses from our mock observations using the basic assumptions, which are usually applied when inferring the dust disk masses and radii.  In particular, we assume that the dust disk size $R_{\rm dust}^{\rm obs}$ is defined by the radial extent, within which 95\% of total flux $F_{\nu}$ is contained. To calculate the dust disk mass $M_{\rm dust}^{\rm obs}$, we follow the usual procedure and use an optically thin approximation  \citep[e.g.,][]{Tobin2020,Kospal2021} 
\begin{equation}
    M_{\rm dust}^{\rm obs} = {d^2 F_{\nu}^{95\%} \over B_{\nu}(T_{\rm d}) \kappa^{\rm asm}_\nu },
\end{equation}
$F_\nu^{95\%}$ is the flux contained within the disk extent defined by $R_{\rm dust}^{\rm obs}$, $B_\nu(T_{\rm d})$ the Planck function at the assumed isothermal dust temperature $T_{\rm d}$, and $\kappa^{\rm asm}_\nu$ the assumed dust absorption opacity at 3~mm (per unit mass of dust) set equal to 1.0~cm$^2$~g$^{-1}$ \citep{Beckwith1990}.  The dust temperature is estimated from the following equation \citep{Tobin2020}
\begin{equation}
    T_{\rm d} = 43~K \left( {L_{\rm tot} \over 1.0 \,\mathrm{L_\odot} } \right)^{0.25},
\end{equation}
where $L_{\rm tot}$ is the total (accretion plus photospheric) luminosity of the star in our model.
We note that when calculating the synthetic disk mass we use the assumed dust temperature $T_{\rm d}$ and opacity $\kappa_{\rm \nu}^{\rm asm}$ rather than those known from our model data ($T_{\rm mp}$ and $\kappa_\nu$). 
Indeed, when deriving disk masses from observations, disk temperature and opacity are often not known and in this case assumptions like above are utilized.


We further compare the synthetic observables with the disk radii and masses derived directly from the spatial distribution of dust in our model. In particular, for the dust disk radius $R_{\rm dust}^{\rm mod}$ we take the radial extent, within which 95\% of the total dust mass is localized. The corresponding dust mass constitutes the mass of the dust disk $M_{\rm dust}^{\rm mod}$.

\begin{figure}
\begin{centering}
\includegraphics[width=\linewidth]{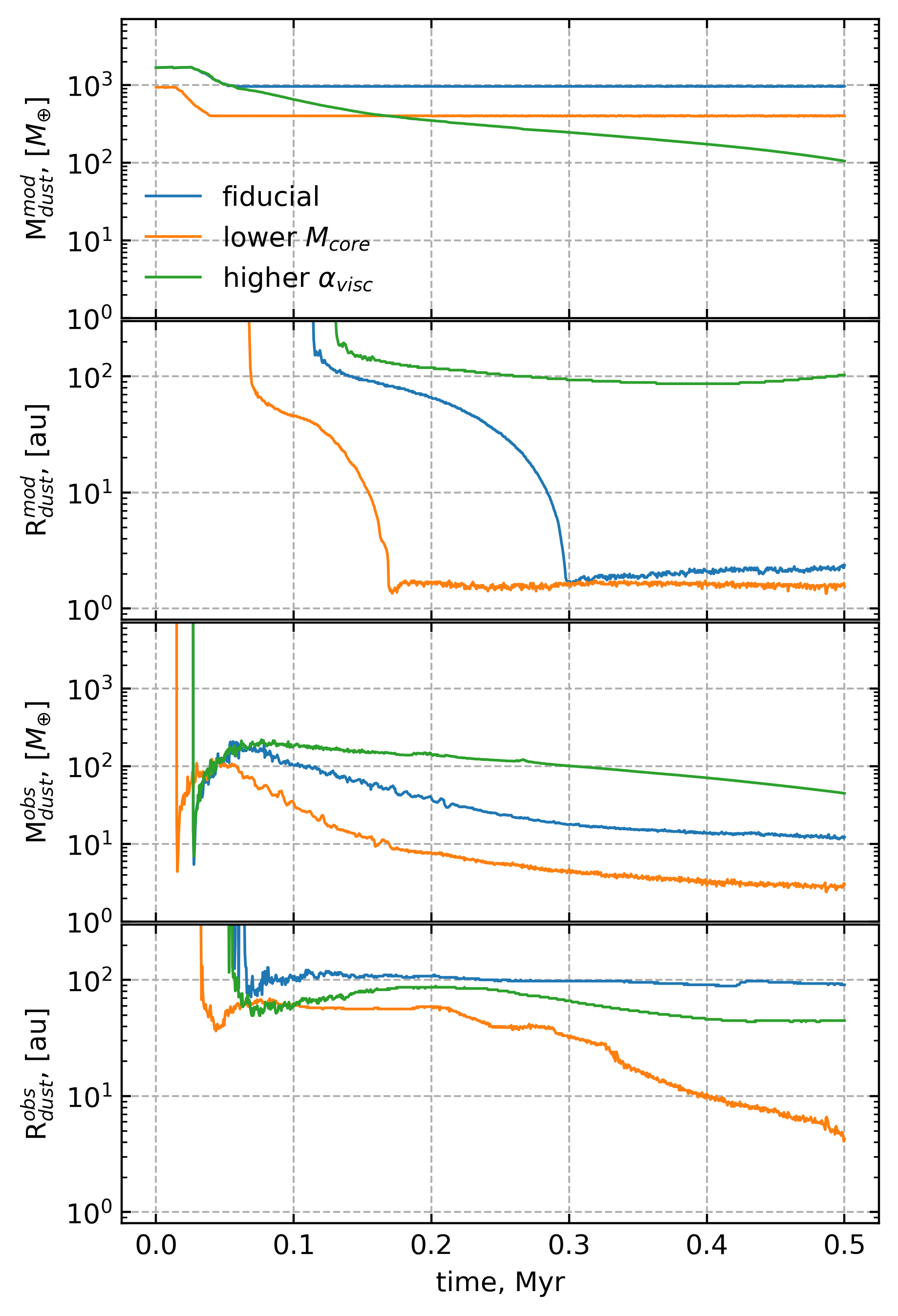} 
\par \end{centering}
\caption{Time evolution of the dust disk masses and radii derived from the model dust distribution (first and second panels, $M_{\rm dust}^{\rm mod}$ and $R_{\rm dust}^{\rm mod}$) and from the mock observations (third and bottom panels, $M_{\rm dust}^{\rm obs}$ and $R_{\rm dust}^{\rm obs}$) in the three models considered. } 
\label{fig:masses-obs-mod}
\end{figure}

In Figure~\ref{fig:masses-obs-mod} we present the synthetic dust disk masses and radii derived using the mock observations and compare them with the corresponding model values as a function of time for the three models considered. Our algorithm for the calculation of $M_{\rm dust}^{\rm mod}$ and $R_{\rm dust}^{\rm mod}$ is applicable to the disk-only stage. In the embedded stage, it may erroneously capture dust in the infalling envelope.  This is the reason why the model disk radii initially start from unrealistically large values. 
Figure~\ref{fig:tot-mass-budget} indicates that the disk-only stage begins after $t\approx 0.1-0.15$~Myr, depending on the model, and this should be taken into account when interpreting the model data. 

 
The first and second panels shows the dust disk mass and radius, $M_{\rm dust}^{\rm mod}$ and $R_{\rm dust}^{\rm mod}$ respectively, directly derived from the model dust distribution. The formation of the GI-induced dead zone in the fiducial and lower $M_{\rm core}$ models effectively traps about half of the total dust mass reservoir, which was initially contained in the corresponding prestellar cloud cores. This effect is also evident in Figure~\ref{fig:tot-mass-budget}. The dust disk radius in these models shrinks with time from about 100~au to just several astronomical units, reflecting inward dust drift and efficient trapping of dust in the dead zone.  On the other hand, the higher $\alpha_{\rm visc}$ model features a gradually declining $M_{\rm dust}^{\rm mod}$ owing to the continuing dust drift across the inner disk regions and through the sink cell. Although the dust mass decreases, the dust disk size in this model evolves slowly with time.

The synthetic dust disk masses $M_{\rm dust}^{\rm obs}$ and radii $R_{\rm dust}^{\rm obs}$ presented in the third and bottom panels of Figure~\ref{fig:masses-obs-mod} show a qualitatively different behavior. 
Most of the dust content in the disks of the fiducial and lower $M_{\rm core}$ models is trapped in a narrow optically thick ring around 1~au  with the optical depth as high as hundreds at 3~mm (see Fig.~\ref{fig:mock-fluxes}). This results in a serious underestimate of the dust disk mass derived from mock observations by about two orders of magnitude. A qualitatively similar effect is seen in the model with higher $\alpha_{\rm visc}$ but of a lesser proportion. The fiducial and lower $M_{\rm core}$ models are in general characterized by much lager radii derived from the mock observations than directly from the model dust distribution. On the contrary, the higher $\alpha_{\rm visc}$ model  features lower $R_{\rm dust}^{\rm obs}$ compared to the corresponding values of $R_{\rm dust}^{\rm mod}$. We conclude that the real and observationally inferred dust disk masses and radii may differ significantly, in agreement with our earlier numerical experiments \citep{2014DunhamVorobyov}. The apparent deficit of dust mass needed to explain the formation of the observed planetary systems, as inferred from observations of Class II disks in particular, reinforces our findings that a substantial dust mass reservoir may be hidden from our view \citep{Manara2018,Miotello2023}

\section{Discussion and model caveats}
\label{Sect:caveats}
The dust pile-up followed by the presumed development of the streaming instability occurs in the GI-controlled disk soon after the disk formation instance.  Planetesimals that may be formed through this process will represent the first building blocks of planets in the terrestrial zone of the disk. These planetesimals may further grow via an oligarchic growth and/or pebble accretion. The early onset of the streaming instability suggested by our numerical simulations is in agreement with the changes in the planet formation paradigm, shifting the onset of planet formation to the Class I and even Class 0 phases \citep{Vorobyov2011,GreavesRice2011,Brogan2015}. 
We note, however, that the onset of planetesimal formation in the dust ring should inevitably change its appearance, as a substantial fraction of dust may be converted to planetesimals. The optical depth and temperature of the corresponding disk region will drop. All these effect we plan to explore self-consistently in a follow-up study. 

Our proposed mechanism for the dust ring formation crucially depends on the existence of a gravitationally unstable phase in the evolution of young protoplanetary disks.   
Many numerical studies have demonstrated that GI can be triggered in sufficiently massive protoplanetary disks, $\ge 0.1 M_\odot$ \citep[see a review by][]{Kratter2016}. The conditions are particularly favorable in the embedded stage of disk evolution, when continual mass loading from the infalling envelope helps to sustain and enhance the GI in the disk \citep{2005VorobyovBasu}. Magnetic fields do not impede the development of GI \citep{2014Machida,2018ZhaoCaselli}.

From the observational point of view, however, GI remains elusive. The direct manifestation of GI -- a spiral pattern -- is indeed observed in several protoplanetary disks \citep{2016Perez,Parker2022}, but its origin is debated and may be caused not only by GI \citep{Meru2017}, but also by an embedded planet \citep{Dong2017}. 

Furthermore, our model may appear to contradict strong dust settling inferred for many protoplanetary disks \citep[e.g.,][]{Rosotti2023}. Indeed, $\alpha_{\rm GI}$ is substantial beyond several astronomical units ($> 10^{-3}$), see Figure~\ref{fig:alpvisc_alpeff}, and this can hinder dust settling towards the midplane owing to substantial gravitoturbulent vertical stirring \citep{Riols2020}.  This contradiction may be lifted twofold. First, we note that $\alpha_{\rm GI}$  determines the efficiency of gravitational torques as a means of mass and angular momentum transport in the disk midplane (see Eq.~\ref{eq:alpha-GI}).  The vertical Reynolds stress tensor, which defines the strength of vertical mixing in a GI-controlled disk, may be weaker than the gravitational stress tensor in the disk midplane \citep{Baehm2021}. The effect of GI is then anisotropic, which may assist dust settling. 
Second, protoplanetary disks with efficient dust settling may be already in the evolution stage that is past the gravitationally unstable phase. Indeed, recent observations of young disks in the Class 0 and I stages found little dust settling \citep{Lin2023}.

In our work, we have considered a limited set of disk models. Our disks are fairly massive ($> 0.1~M_\odot$) and readily support GI, but if disks are systematically less massive than $0.1 M_\odot$,  the GI-induced mechanism of the dead zone formation may not work. Fortunately, recent measurements of disk masses in FU Orionis-type objects, most of which are likely to belong to the Class I stage \citep{Quanz2007,VorobyovBasu2015}, found that half of the sample has massive disks, $\ge 0.1~M_\odot$ \citep{Kospal2021}. This observational finding reinforces the feasibility of the GI-induced mechanism for the formation of dead zones.

We also note that the position of the inner edge of the disk at 0.2~au (radius of the sink cell) does not correspond to the true inner disk edge, which is usually located at several stellar radii. This may affect the location of the dust ring in our models. However, resolving the inner disk edge is only possible in one-dimensional disk models, which cannot self-consistently simulate gravitational instability \citep[e.g.,][]{Gehrig2021}, and is beyond the capacity of multidimensional codes that follow disk formation and evolution on Myr-scales like \simname{FEOSAD}.
In the future works, we will add a possibility of dust trapping at the water snow line \citep{2017Drazkowska} and consider the potentially important effects of magnetic disk winds.

 Finally, we want to comment on the gravity force calculations that were utilized in 
\simname{FEOSAD} (see Appendices~\ref{Sect:gravpot} and \ref{App:test-runs} for details). Many studies of two-dimensional self-gravitating thin disks include a smoothing length $\epsilon$ when calculating the gravitational potential \citep[e.g.,][]{Baruteau2008,Hure2005,Kley2012,2023A&A...675A..96R}. It is often claimed that the introduction of the smoothing length to the gravitational potential is necessary 1) to avoid the problem of singularity and 2)  to better reproduce the three-dimensional potential on the underlying two-dimensional grid. However, as was noted in \citet{BT1987}, the problem of singularity in the context of self-gravitating disks (but not for planets embedded in the disk) can be avoided by calculating analytically  the contribution of the material in the singularity cell to the total gravitational potential.  As we demonstrated on test problems with an analytic solution in Appendix~\ref{Sect:gravpot}, our $\epsilon$-free method is only slightly inferior to the best-choice $\epsilon$-correction method. The accuracy of the latter method, however, is quite sensitive to the proper choice of $\epsilon$ (see Figs.~\ref{fig:compare-grav} and \ref{fig:compare-grav-const}), which is often made proportional to the disk vertical scale height $H$.  There is no universal recipe as to what the coefficient of proportionality between $\epsilon$ and $H$ is to take and different studies advocate different values \citep[e.g.,][]{Hure2005,Baruteau2008, Kley2012}. 
In Appendix~\ref{App:test-runs} we carried out test runs with explicit smoothing of the gravitational potential and found that the disk evolution in the fiducial model (no smoothing) is similar to that obtained with the $\epsilon$-smoothing method proposed in \citet{Baruteau2008}.
We conclude that using the $\epsilon$-free method for computing the gravitational potential in the thin-disk simulations is justified considering all the limitations of the two-dimensional approach in general.

\section{Conclusions}
\label{Sect:conclude}
In this work, we studied in detail a new mechanism of the dead zone formation in the inner regions of protoplanetary disks, which occurs during the initial gravitationally unstable stages of disk evolution if the MRI turbulence is suppressed across the disk extent. We considered the efficiency of dust accumulation in these GI-induced dead zones using the \simname{FEOSAD} code, which computes the formation and long-term evolution of gravitationally unstable gas-dust disks in the thin-disk limit. Our main findings can be summarized as follows.
\begin{itemize}
    \item Gravitationally unstable disks are characterized by a radially varying strength of gravitational instability. The effects of this variation, when quantified in terms of $\alpha_{\rm GI}$, are similar to the classical layered disk model. Namely, a region of low mass and angular momentum transport forms in the inner several astronomical units of the disk, where GI is suppressed. This region is similar in characteristics to the dead zone that usually forms in the layered disk model.

    \item Grown dust that drifts from the outer disk regions efficiently accumulates in the GI-induced dead zone, leading to the formation of a massive dust ring around 1~au. The dust ring is susceptible to the development of the streaming instability. 

    \item The dust ring and the streaming instability occur as early as 15~kyr after the disk formation instance. Hence, this mechanism may form the first generation of planetesimals, which may constitute the first building blocks for planets in the inner terrestrial zone of the disk.

    \item For the GI-induced dead zones and dust rings to form, the MRI has to be suppressed across the disk extent. Increasing $\alpha_{\rm visc}$ due to MRI to $10^{-3}$ results in a  much shallower dead zone, weaker and more diffuse dust ring, and suppression of the streaming instability. We note that MRI suppression in disks with enhanced GI is possible according to three-dimensional sheared-box simulations of \citet{Riols2018}.

    \item In the gravitationally unstable disks the dust masses and radii calculated directly from the model dust distribution and from mock observations following the usual assumptions about the dust temperature, optical depth, and dust opacity differ significantly. In particular, the dust disk masses derived from mock observations are seriously underestimated. The corresponding dust radii may be larger or smaller than the true underlying radial dust distribution.
    
\end{itemize}


\begin{acknowledgements} We are thankful to the anonymous referee for constructive comments and suggestions that helped to improve the manuscript.
This work was supported by the
Ministry of Science and Higher Education of the Russian
Federation (State assignment in the field of scientific activity 2023, GZ0110/23-10-IF).
Simulations were performed on the Vienna Scientific Cluster (VSC) \footnote{\href{https://vsc.ac.at/}{https://vsc.ac.at/}}.
\end{acknowledgements}

\bibliographystyle{aa}
\bibliography{refs}

\begin{appendix}

\section{Steady-state viscous disk model}
\label{app:steady_state}
\begin{figure}
    \centering
    \includegraphics[width=\linewidth]{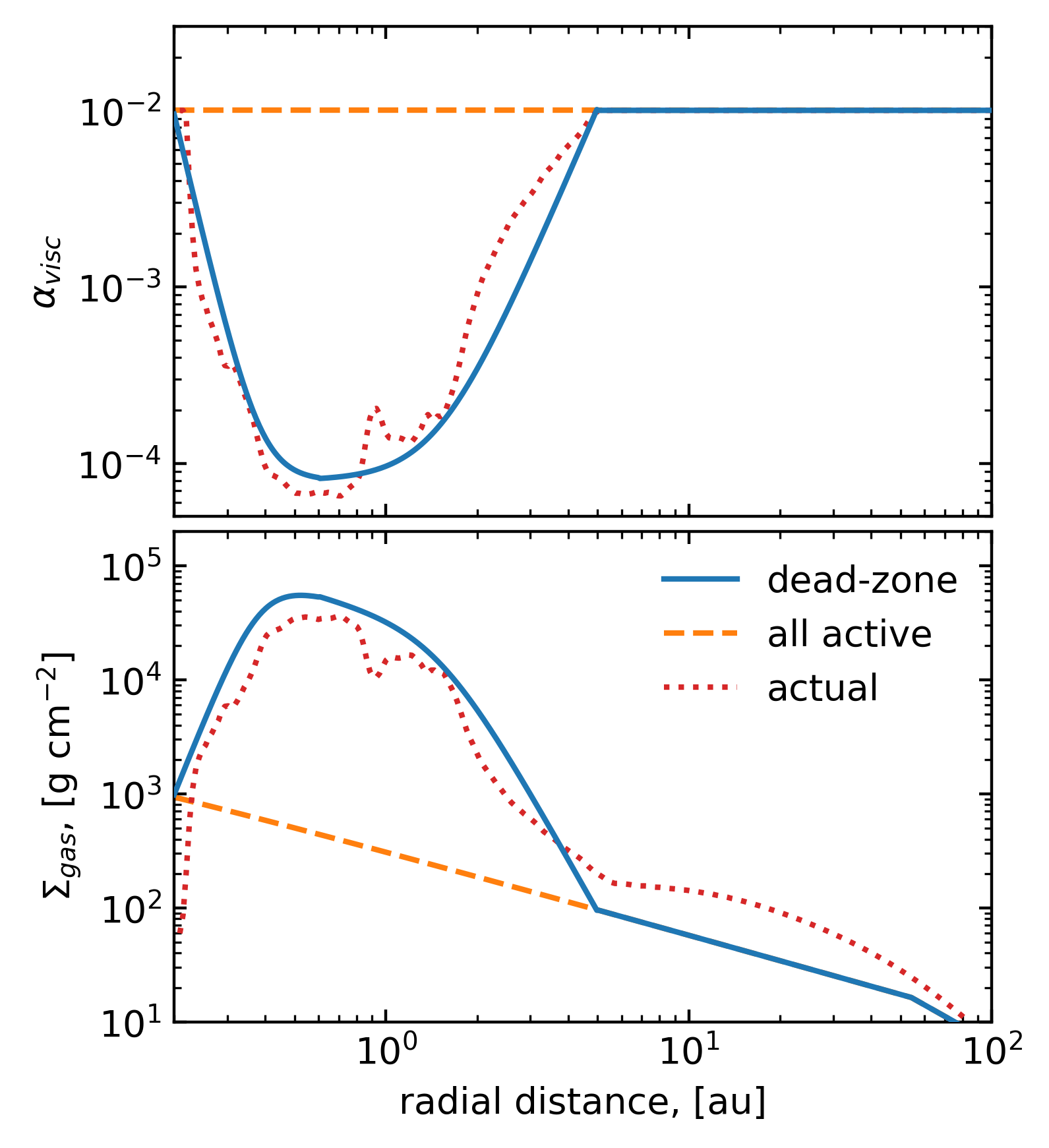}
    \caption{Gas surface density and $\alpha_{\rm visc}$ radial profiles. {{\bf Top}}. The red dotted and blue solid lines show the actual and fitted $\alpha_{\rm visc}$ values in the layered disk model with a dead zone, while the orange dashed line corresponds to an MRI-active disk with a constant $\alpha_{\rm visc}$. {{\bf Bottom}}. The corresponding radial profiles of the gas surface density obtained from the actual hydrodynamic simulation (red dotted line), from solution of the steady-state Eq.~(\ref{eq:1D-sigma}) with radially varying $\alpha_{\rm visc}$ (blue solid line), and  from Eq.~(\ref{eq:1D-sigma}) with a spatially constant $\alpha_{\rm visc}$ (orange dashed line). }
    \label{fig:steady_visk}
\end{figure}

To explain the accumulation of matter in the layered disk model presented in Sect.~\ref{sec:variable_alpha}, it is useful to consider the disk radial structure in the steady-state limit $\partial / \partial t \rightarrow 0$. In this case, for an axisymmetric, geometrically thin but optically thick viscous disk, we can write \citep[see, e.g.,][]{Pringle1981, Hartmann1998, Armitage2022}
\begin{equation}
    \label{eq:1D-sigma}
    \nu \Sigma_{\rm g} = \dfrac{\dot{M}}{3 \pi} \left( 1 - \sqrt{\dfrac{R_{\ast}}{r}}\right),
\end{equation}
where $R_{\ast}$ is the radius of the central star, $\dot{M} = - 2 \pi r \Sigma_{\rm g} v_{\rm r}$ the mass transport rate through the disk, and $v_{\rm r}$ the radial component of gas velocity (negative because of the flow towards the star).

The equation of continuity for gas 
\begin{equation}
    \dfrac{\partial \Sigma_{\rm g}}{\partial t} + \dfrac{1}{r} \dfrac{\partial}{\partial r} \left( r \Sigma_{\rm g} v_{\rm r} \right) = 0 ,
\end{equation}
in the steady-state limit ($\partial \Sigma_{\rm g} / \partial t = 0$) reduces to the following form 
\begin{equation}
    \dfrac{1} {2 \pi} \dfrac{\partial \dot{M}} {\partial r}  = 0. 
\label{eq-dot-M}
\end{equation}
Equation~(\ref{eq-dot-M}) indicates that in a steady-state axisymmetric disk the rate of mass transport across the disk does not depend on the distance to the star $r$. Therefore, for a fixed $\dot{M}$, the radial distribution of the gas surface density in a steady-state disk is exclusively determined by the kinematic viscosity $\nu$.

\begin{table}
    \centering
        \caption{Coefficients of Eq.~(\ref{alpha-visc-fit})}
    \begin{tabular}{c||c|c|c}
       \hline 
       zone, [au]  & A & b & C \\
       \hline 
       \hline 
      $0.2 \leqslant r \leqslant 0.6$ & $7.45 \times 10^{-8}$ & -7.3 & 8 $\times 10^{-5}$ \\
      $0.6 < r \leqslant   5$ & $1.64 \times 10^{-5}$ &    4 & $8 \times 10^{-5}$ \\
       r > 5 & $10^{-2}$ &    0 & 0
    \end{tabular}
    \label{tab:steady_visc}
\end{table}

We now consider an optically thick disk, in which the gas temperature is determined according to the law $T(r) \propto r^{-3/4}$ \citep{Hartmann1998}. With an assumption of hydrostatic equilibrium in the vertical direction, the local scale height of the gaseous disk can be expressed in terms of the local speed of sound, as $H_{\rm g} = c_{\rm s} / \Omega_{\rm K}$. 
We further
set the radial profile of $\alpha_{\rm visc}$ such that there is a dead zone with a suppressed MRI at $0.2 \leqslant r \leqslant 5$~au and the rest of the disk is MRI-active. The $\alpha_{\rm visc}$ value in the active region is fixed at $10^{-2}$, in accordance with the model described in Sect.~\ref{sec:variable_alpha}. 
In the dead zone, we use the $\alpha_{\rm visc}$ profile, which is similar to that found in the layered disk model at $t=375$~kyr. Particularly, the following 
function is used to describe the $\alpha_{\rm visc}$ radial profile in the dead zone
\begin{equation}
 \alpha_{\rm visc} = A \cdot r^{b} + C.
 \label{alpha-visc-fit}
\end{equation}
The dead zone area is divided into 2 parts: the regions of decreasing and increasing $\alpha_{\rm visc}$ with distance $r$. The transition occurs at the point where $b$ changes sign and absolute value. The coefficient $C$ is used to smooth the distribution, and the factor $A$ is chosen so as to eliminate discontinuities at the transition boundaries. The values of coefficients $A$, $b$, and $C$ adopted in the work are given in Table~\ref{tab:steady_visc}.

The top panel of Figure~\ref{fig:steady_visk} displays the fitted $\alpha_{\rm visc}$ (solid blue line) along with the actual $\alpha_{\rm visc}$ values taken from the simulation (red dotted line). The value of $\dot{M}$ is set equal to $7\times 10^{-8}~M_\odot$~yr$^{-1}$, which is consistent with the model accretion rates onto the protostar, found to be in the range of $(7-9)  \times 10^{-8}~M_\odot$~yr$^{-1}$ at the corresponding evolutionary time. For comparison, the orange dashed line represents the case of a fully MPH-active disk. The blue line in the bottom panel shows the surface density profiles of gas calculated according to Equation~(\ref{eq:1D-sigma}) for given temperature and viscosity distributions, while the red dotted curve presents the actual gas surface density obtained in our numerical simulations.  In the model of a fully MPH-active disk, the surface density profile corresponds to a monotonically decreasing function of radial distance $r$. At the same time, in the model with the presence of a dead zone, there is an accumulation of matter, which qualitatively (and quantitatively with a factor of two) agrees with the distribution obtained in hydrodynamic simulations in Sect.~\ref{sec:variable_alpha}.

\section{Gravitational potential calculation}
\label{Sect:gravpot}

The gravitational potential of the disk in the polar coordinates $(r,\phi)$ can be written as
\begin{eqnarray}
\label{App:pot}
   \Phi(r,\phi) & =&  \\ \nonumber
   &-& G \int_{\rm r_{\rm sc}}^{r_{\rm out}} r^\prime dr^\prime
                   \int_0^{2\pi}
                \frac{ \Sigma_{\rm tot} (r^\prime,\phi^\prime) d\phi^\prime}
                     {\sqrt{{r^\prime}^2 + r^2 - 2 r r^\prime
                        \cos(\phi - \phi^\prime) }}  \, ,
\end{eqnarray} 
where $\Sigma$ is the total (gas plus dust) mass. The direct summation of the resulting double sum on the discretized $N\times N$ polar grid is time consuming as it takes $(2N)^4$ operations. As pointed out in \citet{BT1987}, this equation can be transformed to a more manageable form by the following substitution of variables 
$$ u=\ln r; \,\, S = r^{3/2} \Sigma; \,\, V = r^{1/2} \Phi.  $$
The resulting equation reads as
\begin{eqnarray}
\label{App:pot-conv}
   V(u,\phi) & =&  \\ \nonumber
   &-& G \int_{\rm u_{\rm sc}}^{u_{\rm out}} du^\prime \int_0^{2\pi}
                \frac{ 2^{-1/2} S(u^\prime,\phi^\prime) d\phi^\prime}
        {\sqrt{\cosh(u-u^\prime) -
                        \cos(\phi - \phi^\prime) }}  \, ,
\end{eqnarray} 
and can be solved using the convolution theorem, which involves a series of fast Fourier transforms. For details we refer the reader to \citet{BT1987} and \citet{Vorobyov2023-pot}, but note here that this method requires $2N[6\log_2(2N)+1]$ additions and multiplications to be compared to $(2N)^2$ operations for a direct evaluation of Eq~(\ref{App:pot}), which becomes computationally favourable already for $N>16$.  

The usual complication complication with evaluating integrals~(\ref{App:pot}) or (\ref{App:pot-conv}) arises when the primed and non-primed coordinates become equal, because the denominator in this case becomes equal to zero and the sum diverges.   
In many practical applications, the potential is smoothed  by introducing a smoothing length $\epsilon$, so  that Eq.~(\ref{App:pot}) would read as
\begin{eqnarray}
\label{App:pot-smooth}
   \Phi(r,\phi) & =&  \\ \nonumber
   &-& G \int_{\rm r_{\rm sc}}^{r_{\rm out}} r^\prime dr^\prime
                   \int_0^{2\pi}
                \frac{ \Sigma_{\rm tot} (r^\prime,\phi^\prime) d\phi^\prime}
                     {\sqrt{{r^\prime}^2 + r^2 - 2 r r^\prime
                        \cos(\phi - \phi^\prime) +\epsilon^2 }}  \, .
\end{eqnarray} 
After applying the coordinate transformation introduced above, this equation becomes
\begin{eqnarray}
\label{App:pot-conv-reduced}
   V(u,\phi) & =& -G \int_{\rm u_{\rm sc}}^{u_{\rm out}} du^\prime  \\ \nonumber
   &\times& \int_0^{2\pi}
                \frac{ 2^{-1/2} S(u^\prime,\phi^\prime) d\phi^\prime}
        {\sqrt{\cosh(u-u^\prime) -
                        \cos(\phi - \phi^\prime) +0.5\epsilon^2 e^{-(u+u^\prime)} }}  \, ,
\end{eqnarray} 
As noted by \citet{Baruteau2008}, the introduction of the smoothing length breaks the convolution property of the expression in the denominator of Eq.~(\ref{App:pot-conv-reduced}). However, this property can be restored, if a specific form of the $\epsilon$ dependence that is proportional to the disk radius is used, $\epsilon\propto r\propto e^u$. Since the disk vertical scale height $H$ happens to be also proportional to the radial distance in protoplanetary disks, this makes it useful to relate $\epsilon$ to $H$. We note that the introduction of the smoothing length violates the Newton's law of gravity, but it has an advantage of avoiding the problem of singularity. 

This approach has been further developed to better reproduce the three-dimensional potential on the underlying two-dimensional grid by properly adjusting the value of $\epsilon$ \citep[e.g.,][]{2023A&A...675A..96R}. Unfortunately, no universal recipe has been developed and various studies provide different recommendations and prescriptions \citep[see, e.g.,][for a review]{Hure2009}. Even when applied to different simulation environments, the values of $\epsilon$ may differ. For instance \citet{Kley2012}, advocated to set $\epsilon = 0.7 H$  when considering the planet dynamics in the disk, but choose $\epsilon=1.2 H$ when simulating a self-gravitating disk. The problem of non-convergence in the value of $\epsilon$ may be inherit to this method because it depends on the subtleties of the local three-dimensional gas distributions, which change from model to model, and throughout the disk evolution sequence. 

Considering the uncertainties with the smoothing length approach, it is important to note that the use of the smoothing factor can be avoided altogether when evaluating Eq.~(\ref{App:pot-conv}) for self-gravitating disks. As noted by \citet{BT1987}, the contribution of the material in the singularity cell to the total gravitational potential can be evaluated if we assume $S(u^\prime,\phi^\prime)=\mathrm{const}$ and approximate $\cosh(u-u^\prime) - \cos(\phi-\phi^\prime)$ as $0.5(u-u^\prime)^2 - 0.5 (\phi -\phi^\prime)^2$. The resulting contribution to the reduced potential then reads
\begin{eqnarray}
    V(0,0)  = -2\,G\,S \left[ {1 \over \Delta \phi}\sinh^{-1}\left( {\Delta \phi \over \Delta u}\right) + {1\over \Delta u}\sinh^{-1}\left( {\Delta u \over \Delta \phi} \right) \right],  
    \label{app:sing-polar}
\end{eqnarray}
where $\Delta u$ and $\Delta \phi$ are the cell sizes in the $u$- and $\phi$-coordinate directions, and $S$ is the reduced surface density in this cell. This method can also be extended to two-dimensional Cartesian grids, in which case the gravitational  potential in the singularity cell is evaluated as
\begin{eqnarray}
    \Phi(0,0)  = -2\,G\,\Sigma \left[ \Delta x\sinh^{-1}\left( {\Delta y \over \Delta x}\right) + \Delta y \sinh^{-1}\left( {\Delta x \over \Delta y} \right) \right], 
    \label{App:sing-cart}
\end{eqnarray}
where $\Delta x$ and $\Delta y$ are the the corresponding cell sizes on the Cartesian mesh.
For more complex cases of three-dimensional potentials, with and without an assumption of the constant density inside the singularity cell, we refer the reader to \citet{Macmillan1958} and \citet{Hahn2020}.


We now consider in more detail the $\epsilon$-free method outlined above. Although we avoid introducing an explicit $\epsilon$-factor in Eq.~\ref{App:pot-conv}, we still smooth the potential over the size of the singularity cell by means of the simplified calculation of the potential in this cell. The assumption of $S=\mathrm{const}$ translates to $\Sigma\propto r^{-3/2}$, which is in reasonable agreement with the expected surface density profile in gravitationally unstable disks that are self-regulated around Toomre $Q=1$ \citep{Rice2009,2018VorobyovAkimkin}. 
Furthermore, since the cell size on our logarithmically spaced grid scales linearly with distance $r$, the implicit smoothing that is inherent to our method also scales near linearly with $H$, as advocated by, e.g., \citet{Baruteau2008} and \citet{Kley2012}. Indeed, $\Delta r / r = 0.025$ for our grid, while $H/r$ is a weakly varying function of radius and takes values of 0.05 at 1.0~au and 0.1 at 100~au (for our fiducial model at $t=0.5$~Myr). This means that $\Delta r \propto 0.25-0.5 H$ in our models and the inherent smoothing is proportional to the vertical scale height, as is often assumed in the explicit $\epsilon$-correction models \citep[e.g.,][]{Baruteau2008,Kley2012}.   

\subsection{Exponentially declining disk}
\begin{figure}
    \centering
    \includegraphics[width=\linewidth]{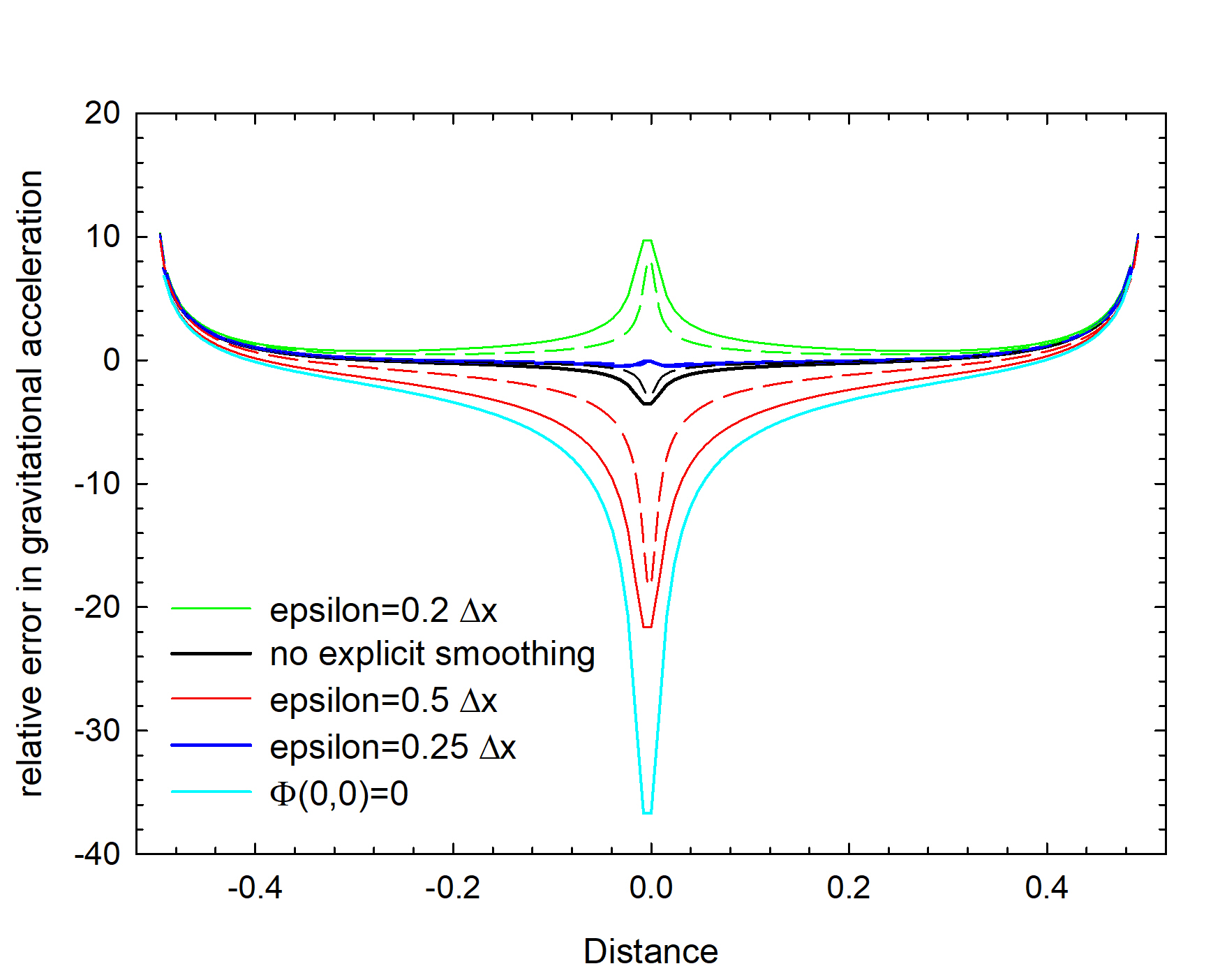}
    \caption{Relative error to the gravitational acceleration. The curves show solutions with and without the $\epsilon$-correction as indicated in the legend. The solid and dashed styles correspond to numerical resolutions of $128\times128$ and $256\times256$ grid zones, respectively.  }
    \label{fig:compare-grav}
\end{figure}

Now, we proceed with analytic test problems.
Figure~\ref{fig:compare-grav} compares the numerically derived gravitational accelerations with the analytic solution for a disk with an exponential surface density distribution of the form $\Sigma=\Sigma_0 \exp(-r/r_0)$. The analytic solution in this case is given by the following equation \citep{BT1987}
\begin{equation}
    \Phi(r)=-\pi \, G \, \Sigma_0 r \left(I_0(y)K_1(y)-I_1(y) K_0(y) \right),
\end{equation}
where $y=r/(2r_0)$, and $I_n$ and $K_n$ are the modified Bessel functions of the first and second kind. A similar test case was used by \citet{Hure2005}. We choose $\Sigma_0=10$ and $r_0=0.1$. The size of the disk and the gravitational constant are set equal to unity. A square grid of $N\times N$ cells is generated, where $N$ takes values of 128 or 256. We note that we intentionally use the Cartesian grid and not the polar grid for this test problem, because for the polar grid we would have to carve out a gap in the innermost disk to avoid the divergence of $u=\ln r$ at the center of the polar coordinates. The analytic solution, however, does not take that gap into account.
In all aspects, the potential solver on the Cartesian mesh is similar to that on the polar mesh, except that it uses Eq.~(\ref{App:sing-cart}) rather than Eq~(\ref{app:sing-polar}) to account for the singularity when the primed and non-primed indices coincide.

The relative errors defined as $(g_x - g_{\rm x,a}) / g_{\rm x,a}$, where $g_x$ and $g_{\rm x,a}$ are the numerical and analytic accelerations along the $x$-axis, respectively, indicate that the best solution is found for the model with a smoothing length $\epsilon=0.25 \Delta x$, where $\Delta x$ is the size of the grid cell. However, the solution that employed smoothing lengths quickly deteriorates as $\epsilon$ deviates from the best value, signaling the strong sensitivity of the method to the proper choice of smoothing. Our method that does not use explicit smoothing yields a fairly good agreement with the analytic solution and is only slightly inferior to the best-fit case of the $\epsilon$-approach. 
The trend shown in Fig.~\ref{fig:compare-grav} is remarkably independent of the numerical resolution, and only the accuracy of both methods improves, as we increase the number of grid cells. To emphasize the importance of calculating the contribution of the material in the singularity cell to the total potential in our method, we artificial set $\Phi(0,0)=0$ (see Eq.~\ref{App:sing-cart}). The resulting relative error is shown by the cyan line. Clearly, the correct calculation of $\Phi(0,0)$ is crucial for our method. We note that the numerical solutions in all methods diverge near the disk outer edge because the analytic solution is obtained for a disk of infinite size. 

\subsection{Constant density disk}
Here, we compare the numerical solutions of the gravitational acceleration in models with and without explicit smoothing of the potential using a disk with constant surface density $\Sigma$ and fixed inner and outer radii, $r_{\rm in}$ and $r_{\rm out}$, respectively. The solution for such a  disk can be found analytically \citep[e.g.][]{Durand1964}, which has been used as a test case for gravitational potential models in \citep{Baruteau2008} and \citet{2005A&A...433L..37P}.
The analytic expression for the gravitational acceleration in the disk plane $g_r(r)$ is 
\begin{equation}
    g_r(r) = 4G\Sigma \left( \frac{E(r/r_{\rm out}) + K(r/r_{\rm out})}{r/r_{\rm out}} + K(r_{\rm in}/r) - E(r_{\rm in}/r) \right), 
\end{equation}
where $K$ and $E$ are the complete elliptic integrals of the first and  second kinds, respectively. The expression applies to $r_{\rm in} < r < r_{\rm out}$.

\begin{figure}
    \centering
    \includegraphics[width=\linewidth]{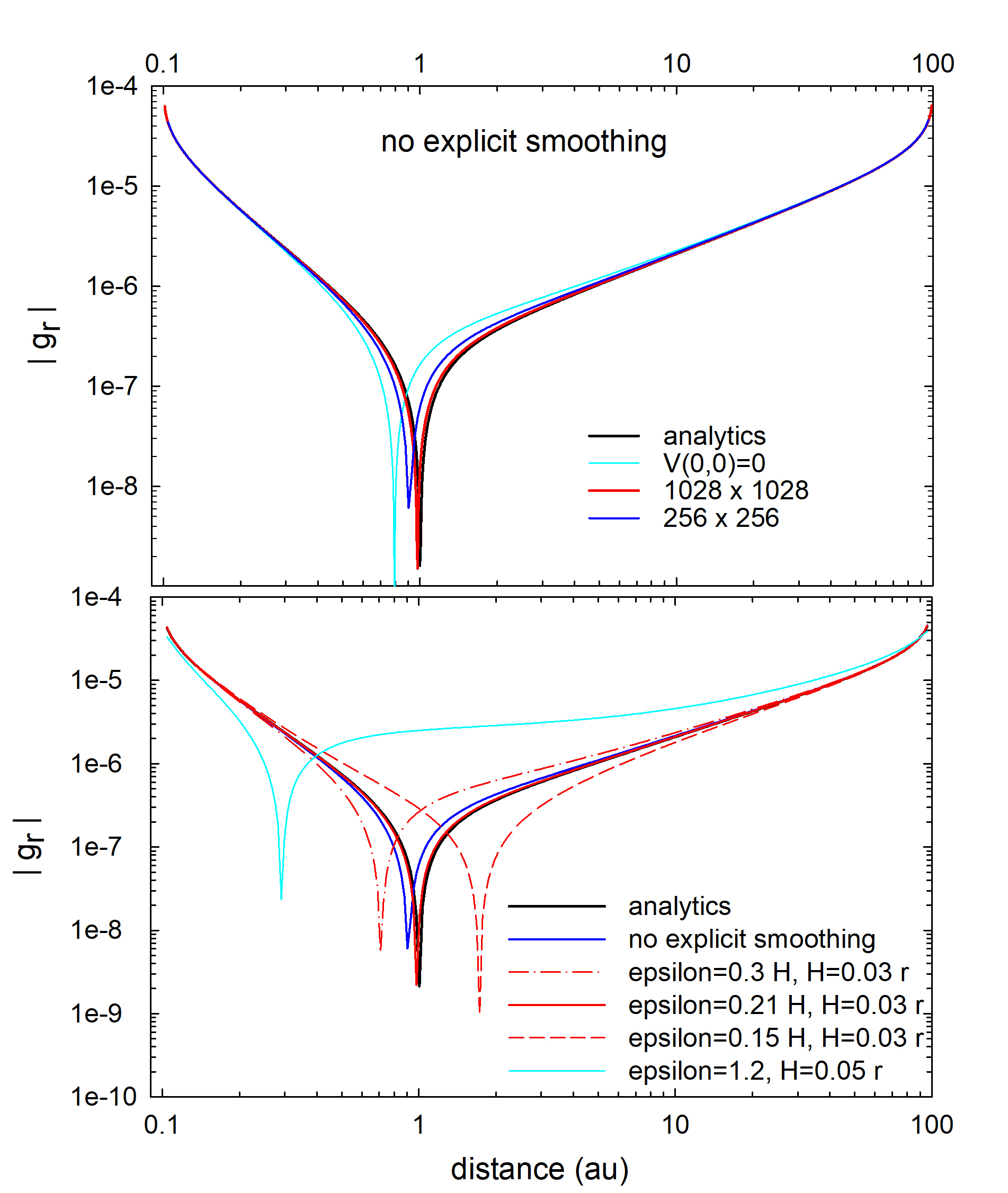}
    \caption{Gravitational acceleration of a constant density disk in models without and with gravitational potential smoothing. The top panel presents the comparison of our method without explicit smoothing with the analytic solution for different number of grid cells and also without considering the contribution of the singularity cell ($V(0,0)=0$). The bottom panel plots the numerical solutions with explicit smoothing for different choices of the $\epsilon$-value. }
    \label{fig:compare-grav-const}
\end{figure}

Figure~\ref{fig:compare-grav-const} presents the results of our numerical experiments. This time, we use physical units and set $r_{\rm in}=0.1$~au, $r_{\rm out}=100$~au, and $\Sigma=100$~g~cm$^{-2}$. The polar grid ($r,\phi$) with $N\times N$ grid cells is initialized.  The gravitational acceleration of the $\Sigma=\mathrm{const}$ disk with a central hole changes sign near the inner edge of the disk and using the relative error is not appropriate in this case. Therefore, we plot the absolute values of the gravitational acceleration $g_{\rm r}$ along the $r$-coordinate direction. 

The top panel shows our method without explicit smoothing of the potential. The numerical solution converges towards the analytic one as the numerical resolution increases from $N=256$ to $N=1028$ grid cells per coordinate direction. If we neglect the contribution of the material in the singularity cell to the total gravitational potential, setting $V(0,0)=0$ (see Eq.~\ref{app:sing-polar}), the solution expectedly deteriorates.
The bottom panels displays the comparison of our method with that using explicit smoothing of the potential. Different combinations of $\epsilon$ are considered according to suggestions put forward in \citet{Baruteau2008}, namely, $\epsilon=0.3 H$, and in \citet{Kley2012}, namely, $\epsilon=1.2 H$. The proportionality between the disk scale height $H$ and the radial distance $r$ is chosen as typical of the fiducial model, $H=0.03-0.05 r$. Clearly, the choice of $\epsilon=1.2 H$ poorly fits the analytic solution. Smaller smoothing lengths can fit the analytic solution better, with $\epsilon=0.21 H$ providing almost a perfect fit. However, small deviations from the best-fit value of $\epsilon$ quickly deteriorate the solution. Our model, though being slightly inferior to the best-fit $\epsilon$-smoothing method, is nevertheless free from uncertainties in choosing the proper value of $\epsilon$.
 This conclusion, however, applies to razor-thin disks. In the case of disks with finite vertical structure, its validity has to be proven by comparing the vertically averaged gravity force of three-dimensional density distributions with the gravity force obtained in the two-dimensional approach with and without the smoothing length \citep{Hure2009,2011A&A...531A..36H,2015MNRAS.447.1866H}. Such a focused study lies beyond this work.

\section{Comparison of disk evolution with and without $\epsilon$-smoothing}
\label{App:test-runs}

In this section, we compare the fiducial model with two additional models that employ the explicit smoothing of the gravitational potential but otherwise are identical. Because numerical simulations of the entire considered disk evolution period are computationally costly, we continue simulations from $t=0.5$~Myr but with different approaches to calculating the gravitational potential. In particular, we consider the smoothing parameters suggested in \citet{Baruteau2008}, namely, we set $\epsilon=0.3 H$ and $H=0.03 r$. In the second model, we set $\epsilon=1.2 H$ and $H=0.05 r$, as advocated by \citet{Kley2012}. In both cases, the adopted relation between the gas scale height $H$ and radial distance $r$ is close to what is found in the fiducial model inside 100~au.

\begin{figure}
    \centering
    \includegraphics[width=\linewidth]{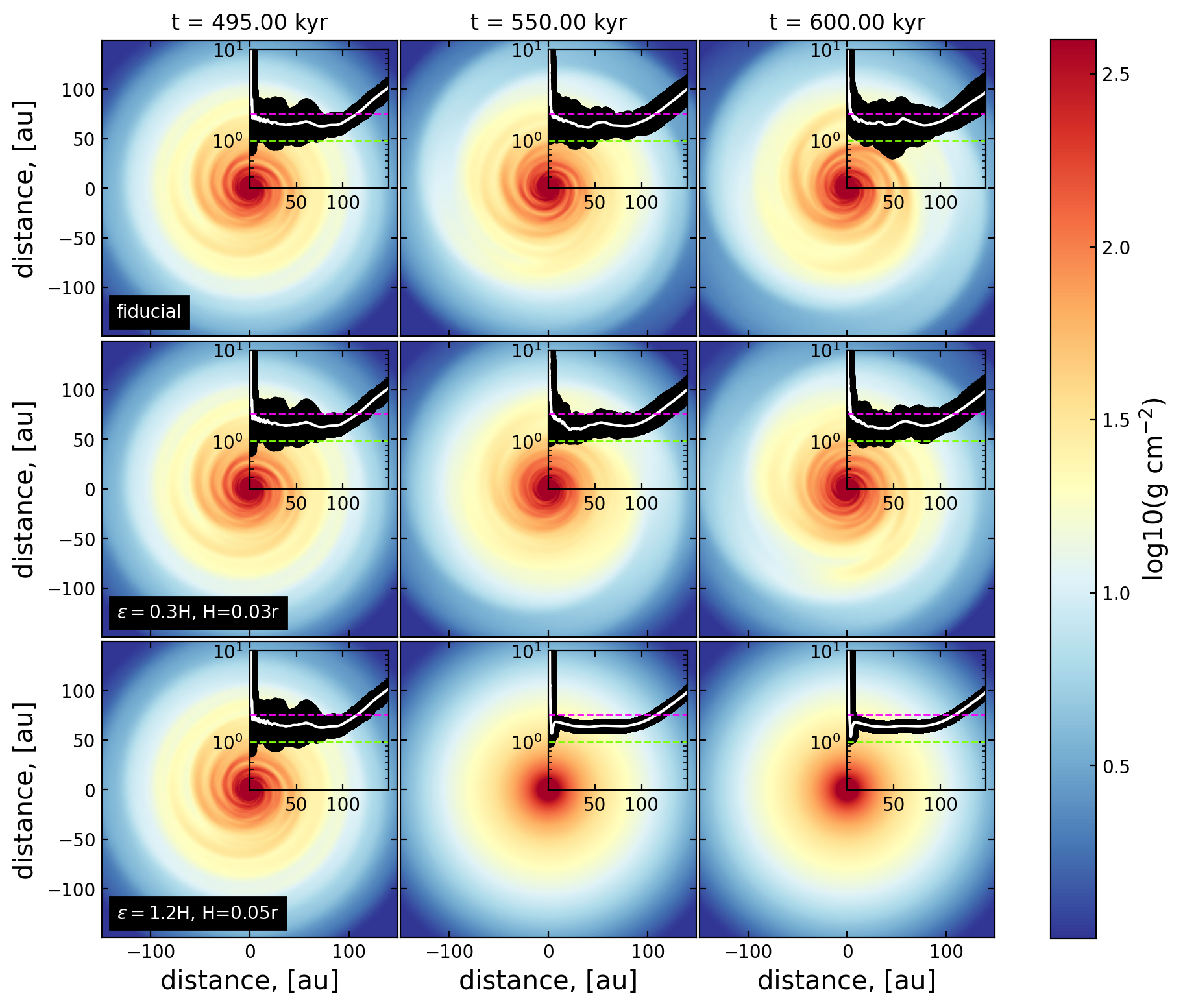}
    \caption{Gas surface density in the fiducial model (top row) and in the models with different smoothing of the gravitational potential (middle and bottom rows). The columns from left to right correspond to evolution times: $t=495$~kyr, $t=550$~kyr, and $t=600$~kyr
     The insets in each of the panels display the Toomre $Q$-parameter as a function of radial distance. The dashed pink and green lines correspond to $Q=2$ and $Q=1$ for convenience. The white lines show the azimuthally averaged values.}
    \label{fig:compare-disk}
\end{figure}

Figure~\ref{fig:compare-disk} presents three snapshots of the gas disk in the considered models at three evolution times: 0.495~Myr, 0.55~Myr, and 0.6~Myr. In addition, the insets show the radial distribution of the Toomre $Q$-parameter in each model and each considered time instance. All values along the azimuth at a given distance $r$ are plotted. Clearly, our fiducial model without explicit smoothing and the model with $\epsilon$-smoothing as suggested by \citet{Baruteau2008} show similar behavior. In both the Toomre parameter is mostly confined in the $Q=1-2$ limits and the disks show a weak spiral pattern, as expected from the gravitational stability analysis \citep{Toomre1964,Polyachenko1997}. 

However, the model with stronger smoothing ($\epsilon=1.2 H$ and $H=0.05 r$) deviates notably and quickly arrives at the gas distribution that is almost axisymmteric.  Curiously, the Toomre parameter stays in similar limits ($Q\sim 1-2$), only featuring a narrower spread. In particular, the azimuthally average $Q$-value at $t=0.55$~Myr and 40~au is 1.47. The disk gravitational stability to local non-axisymmetric perturbations should occur at $Q>\sqrt{3}\approx 1.73$ \citep{Polyachenko1997}, but this model features $Q$-parameters that are lower than the threshold value. We also checked the ratio of the disk to stellar mass and it is greater than 0.1, a value that is often referred to as a threshold for the development of GI in full three-dimensional simulations \citep{Cossins2009,Kratter2016}. This line of evidence indicates that the model with $\epsilon=1.2 H$ and $H=0.05 r$ should be gravitationally unstable but strong smoothing of the gravitational potential appears to prevent its growth.

\begin{figure}
    \centering
    \includegraphics[width=\linewidth]{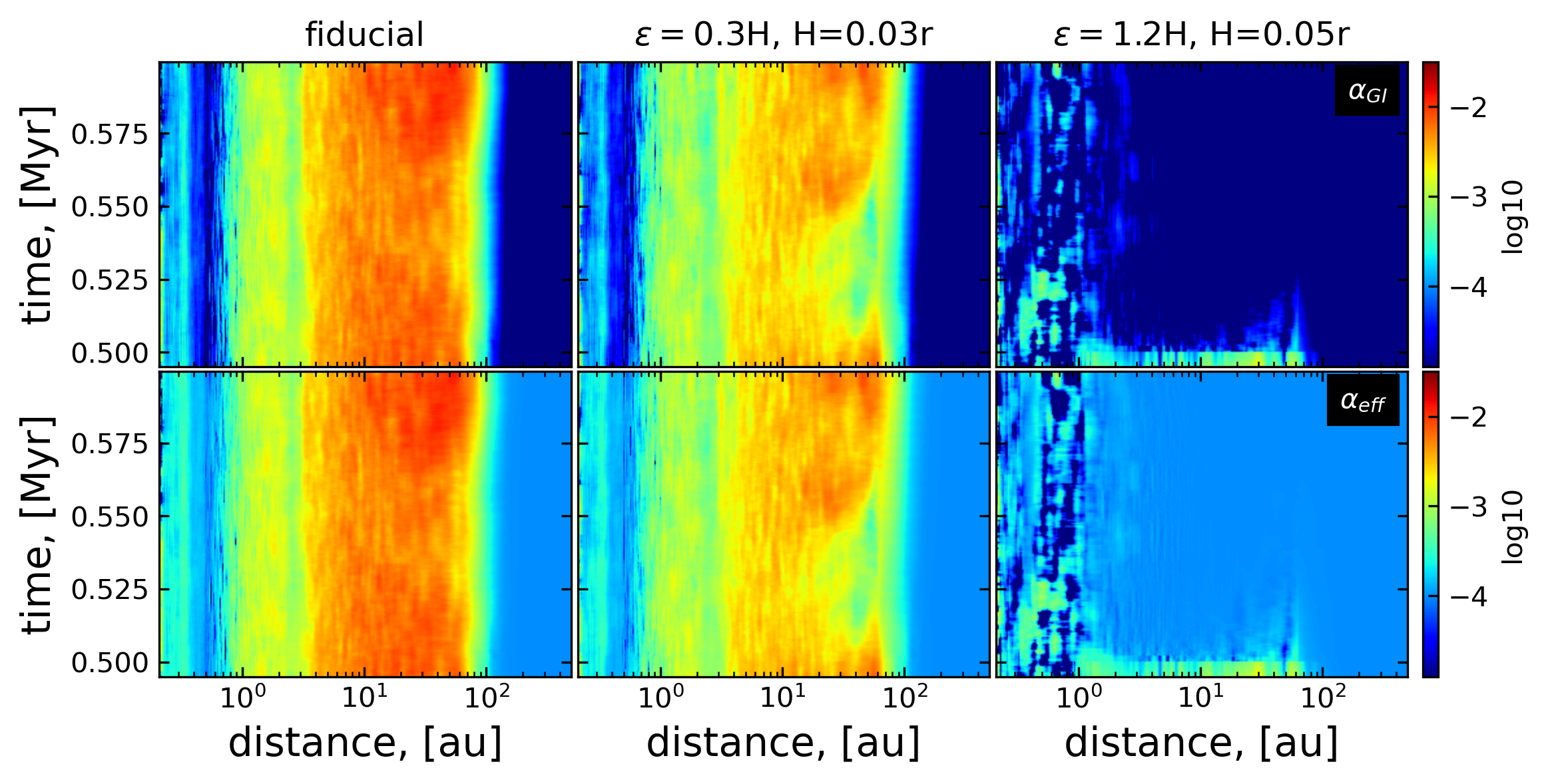}
    \caption{Time evolution of the azimuthally averaged $\alpha$ parameters. The top  and bottom panels show $\alpha_{\rm GI}$ and $\alpha_{\rm eff}$, respectively. Columns from left to right correspond to the fiducial model and to the models with different smoothing of the gravitational potential. }
    \label{fig:compare-alpha}
\end{figure}

Figure~\ref{fig:compare-alpha} presents the space-time plots of $\alpha_{\rm GI}$ and $\alpha_{\rm eff}$ in the three models considered. The spatial and temporal behavior of both parameters are similar in the fiducial model and in the model with weaker smoothing ($\epsilon=0.3 H$ and $H=0.03 r$). The latter model may feature slightly lower $\alpha$-values, but the strong spatial gradient is present in both model. The model with stronger smoothing  ($\epsilon=1.2 H$ and $H=0.05 r$) has much smaller $\alpha$ values. The entire disk in this model is formally a global dead zone from the point of view of the $\alpha$-parameter. 

\begin{figure}
    \centering
    \includegraphics[width=\linewidth]{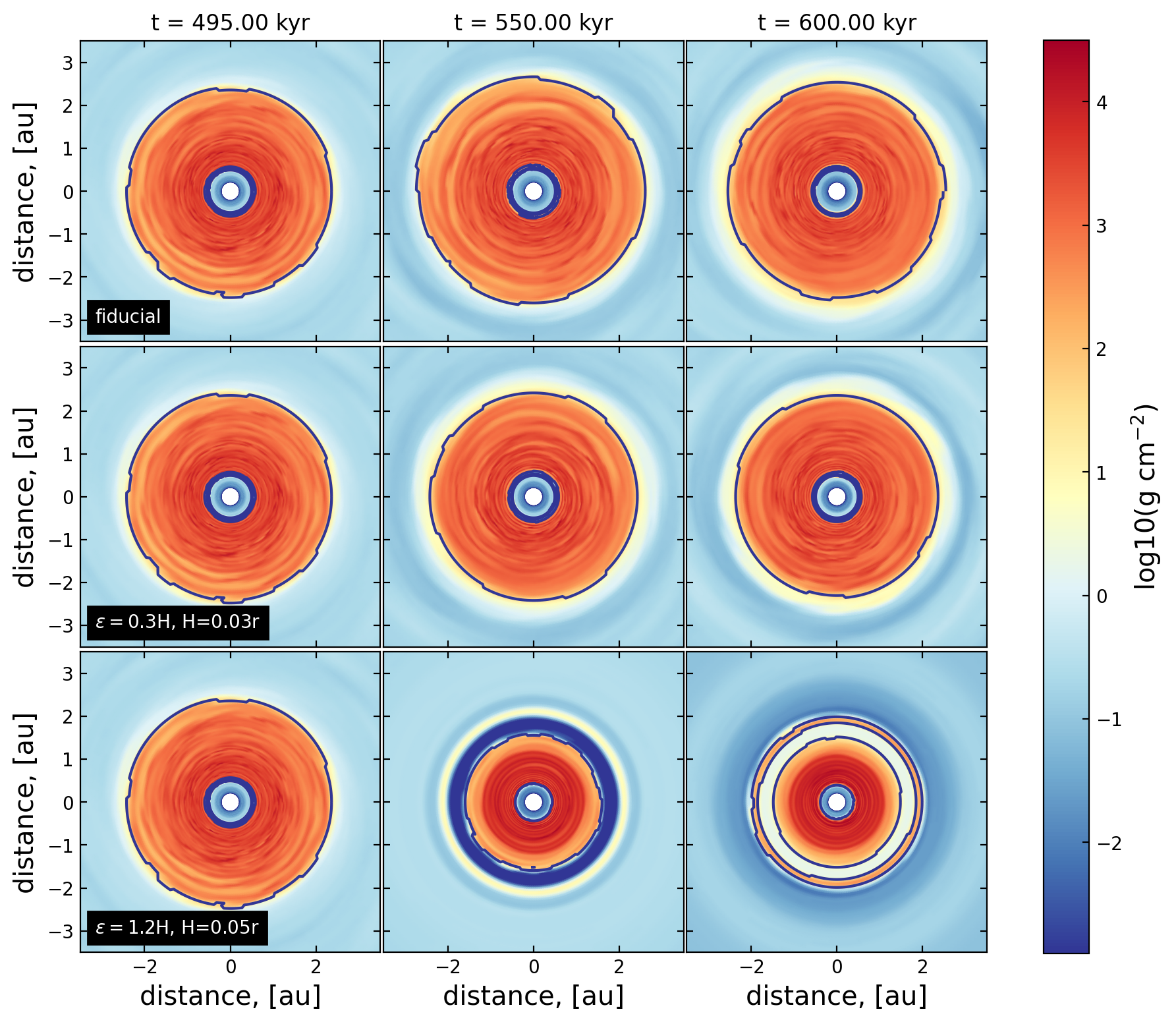}
    \caption{Time evolution of the dust ring in models without and with explicit smoothing of the gravitational potential. The rows from top to bottom show the dust ring in the fiducial model without explicit smoothing, and in models with weaker and stronger smoothing. The black curves outline the regions prone to develop the streaming instability according to Eqs.~(\ref{eq:SI_cond_1}) and (\ref{eq:SI_cond_2}).}
    \label{fig:compare-rings}
\end{figure}

Finally, Figure~\ref{fig:compare-rings} presents the zoomed-in view on the disk inner region encompassing the dust ring in each model considered. The regions that are prone to develop the streaming instability are also shown. The evolution of the dust ring and the SI-prone  disk regions are similar in the fiducial model and in the model with weaker smoothing ($\epsilon=0.3 H$ and $H=0.03 r$). The evolution of the dust ring in the model with stronger smoothing ($\epsilon=1.2 H$ and $H=0.05 r$) deviates notably from the other two models. The dust ring shrinks with time, although it is still susceptible to the streaming instability. 

With all these tests performed, we conclude that our method of calculating the gravitational potential is closest to that proposed in \citet{Baruteau2008}.
We note here that the use of the smoothing factor is often considered as a means of better reproducing the 
three-dimensional potential of a self-gravitating disk when projected on the two-dimensional grid.
It is, however, not clear if using correction factors that modify the Newton's law of gravity can
provide a universal solution to this problem. We leave a more detailed consideration of our method for a future focused study, which will compare realistic nonaxisymmetric potentials on two- and three-dimensional grids.

\end{appendix}

\end{document}